\documentclass[aps,prb,amsmath,amssymb,twocolumn,superscriptaddress,floatfix, longbibliography]{revtex4-2}
\pdfoutput=1

\usepackage{graphicx}
\usepackage{subfigure}
\usepackage{booktabs}
\usepackage{float}
\usepackage{romanbar}
\usepackage[pdftex,colorlinks=true,hypertexnames=false]{hyperref}
\hypersetup{pdfpagemode=UseNone,pdfpagelayout=OneColumn,pdfstartview=FitH}
\begin{document}

\title[]{Contrasting magnetic behavior in \texorpdfstring{MnSc$_2X_4$ ($X$ = S, Se)}
{MnSc2X4} spinel compounds investigated by magnetoelastic studies}

\author{J. Grumbach}
\email{E-Mail: Justus.Grumbach@tu-dresden.de}
\affiliation{Institut für Festk\"orper- und Materialphysik, Technische Universit\"at Dresden, 01062 Dresden, Germany.}

\author{J. Sourd}
\affiliation{Hochfeld-Magnetlabor Dresden (HLD-EMFL) and W\"urzburg-Dresden Cluster of Excellence ct.qmat, Helmholtz-Zentrum Dresden-Rossendorf, 01328 Dresden, Germany.}

\author{M. Deeb}
\affiliation{Institut für Festk\"orper- und Materialphysik, Technische Universit\"at Dresden, 01062 Dresden, Germany.}

\author{A. Miyata}
\affiliation{Hochfeld-Magnetlabor Dresden (HLD-EMFL) and W\"urzburg-Dresden Cluster of Excellence ct.qmat, Helmholtz-Zentrum Dresden-Rossendorf, 01328 Dresden, Germany.}
\affiliation{Institute of Solid State Physics, University of Tokyo, Kashiwa, Chiba, 277-8581, Japan}

\author{H. Suwa}
\affiliation{Department of Physics, University of Tokyo, Tokyo 113-0033, Japan}

\author{T. Gottschall}
\affiliation{Hochfeld-Magnetlabor Dresden (HLD-EMFL) and W\"urzburg-Dresden Cluster of Excellence ct.qmat, Helmholtz-Zentrum Dresden-Rossendorf, 01328 Dresden, Germany.}

\author{A. Hauspurg}
\affiliation{Hochfeld-Magnetlabor Dresden (HLD-EMFL) and W\"urzburg-Dresden Cluster of Excellence ct.qmat, Helmholtz-Zentrum Dresden-Rossendorf, 01328 Dresden, Germany.}

\author{S. Chattopadhyay}
\affiliation{Hochfeld-Magnetlabor Dresden (HLD-EMFL) and W\"urzburg-Dresden Cluster of Excellence ct.qmat, Helmholtz-Zentrum Dresden-Rossendorf, 01328 Dresden, Germany.}
\affiliation{UGC-DAE Consortium for Scientific Research Mumbai Centre, 246-C CFB, BARC Campus, Mumbai 400085, India.}

\author{M. Rotter}
\affiliation{$McPhase~Project$, 01159 Dresden, Germany.}

\author{S. Granovsky}
\affiliation{Institut für Festk\"orper- und Materialphysik, Technische Universit\"at Dresden, 01062 Dresden, Germany.}

\author{L. Prodan}
\affiliation{Experimentalphysik V, Center for Electronic Correlations and Magnetism,
Institute of Physics, University of Augsburg, 86159 Augsburg, Germany.}
\affiliation{Institute of Applied Physics, Moldova State University, 2028 Chisinau, R. Moldova.}

\author{V. Tsurkan}
\affiliation{Experimentalphysik V, Center for Electronic Correlations and Magnetism,
Institute of Physics, University of Augsburg, 86159 Augsburg, Germany.}
\affiliation{Institute of Applied Physics, Moldova State University, 2028 Chisinau, R. Moldova.}

\author{S. Zherlitsyn}
\affiliation{Hochfeld-Magnetlabor Dresden (HLD-EMFL) and W\"urzburg-Dresden Cluster of Excellence ct.qmat, Helmholtz-Zentrum Dresden-Rossendorf, 01328 Dresden, Germany.}

\author{M. Doerr}
\affiliation{Institut für Festk\"orper- und Materialphysik, Technische Universit\"at Dresden, 01062 Dresden, Germany.}

\author{J. Wosnitza}
\affiliation{Institut für Festk\"orper- und Materialphysik, Technische Universit\"at Dresden, 01062 Dresden, Germany.}
\affiliation{Hochfeld-Magnetlabor Dresden (HLD-EMFL) and W\"urzburg-Dresden Cluster of Excellence ct.qmat, Helmholtz-Zentrum Dresden-Rossendorf, 01328 Dresden, Germany.}

\begin{abstract}
The spinel compounds MnSc$_2X_4$ are highly frustrated and candidate materials for
vortex-like 3\textit{\textbf{q}} magnetic states, such as skyrmions, with propagation
vectors in the [111] plane. Because of the strong magnetoelastic coupling, we could
extract a refined magnetic ($H,T$) phase diagram for MnSc$_2$S$_4$ from ultrasound
and dilatometry measurements. We found a variety of magnetic phases, including the
skyrmion phase, which is stable down to lowest temperatures.
In comparison, we investigated MnSc$_2$Se$_4$, having a larger distance between
the magnetic Mn$^{3+}$ ions using the same methods. Unlike in MnSc$_2$S$_4$, we
found no skyrmion phase and overall a lack of sharp anomalies indicative of phase
transitions, neither in dilatometry and ultrasound nor in specific heat and
ac-susceptibility data.
Motivated by our findings, we performed model calculations, which
reproduced the experimentally observed magnetostriction and specific-heat
results reasonably well.
\end{abstract}

\maketitle

\section{Introduction}
The class of spinel compounds containing Mn$^{3+}$ ions is known to show complex
magnetism such as for example spin-spiral or helical ground states, skyrmion phases,
or spin-liquid behavior. Most of these phenomena are driven by asymmetric exchange
or frustration effects \cite{tsurkan21}. Moreover, spinels get attraction because
of their application potential in spintronics or as multiferroic materials. It
should be mentioned that the variation of ions in the usual $AB_2X_4$ structure,
where $A$ and $B$ are usually $3d$ elements and $X$ is a chalcogenide, can
significantly change the magnetism. Therefore, differences are also expected for
the chalcogenides MnSc$_2X_4$ with $X$ = S and Se.

These compounds crystallize in a structure composed of Mn$X_6$ tetrahedra and
Sc$X_4$ octahedra. The chalcogenide corners are shared and a combination of a
Mn-based diamond structure and a Sc-based pyrochlore structure forms the crystal.
Depending on which of the two ionic sites is magnetic, spinels exhibit two types
of magnetic frustration. (i) If the pyrochlore site is magnetic, nearest neighbors
form tetrahedral networks. Therefore, frustration is induced from
nearest-neighbor interaction. An example for such a case is ZnCr$_2$Se$_4$
\cite{cameron_2016, Zajdel_2017, Inosov_2020}. (ii) With magnetic ions on the
diamond site, such as for MnSc$_2X_4$, the nearest-neighbor interactions $J_1$
are not frustrated, whereas the next-nearest neighbors, with interactions $J_2$,
form frustrated triangles in the (111) plane (black triangles in Fig.\ \ref{fig:vesta}).
For $J_2/J_1 > 1/8$, frustration is strong enough to form several types of
helical as well as spiral spin-liquid ground states \cite{bergman_2007, Buessen_2018}.
This condition is fulfilled by both MnSc$_2X_4$ compounds \cite{Guratinder_2022},
so that a variety of magnetic states is present due to frustration.

Neutron-scattering experiments suggest antiferromagnetic skyrmion lattices for
both compounds \cite{gao_2017, gao_2020, rosales, Guratinder_2022}. The twisting
mechanism responsible for the skyrmion texture is induced by frustration, since
spinels are centrosymmetric crystals without Dzyaloshinskii-Moriya interactions.
In this case, skyrmion phases may appear. The size of such skyrmions is
generally small \cite{Kurumaji_2019, Khanh_2020}.

In zero field below 1.5 K, MnSc$_2$S$_4$ shows a commensurable helical magnetic
structure with a propagation vector {\textit{\textbf{q}}} = (3/4, 3/4, 0).
Modifications of this magnetic structure can be explained by structural lattice
changes \cite{krimmel_2006}. Also, theoretical concepts demonstrated the
connection between crystal structure, spin orientation, and frustration in
spinel compounds \cite{bergman_2007,iqbal_2018}. In particular, the strong
spin-lattice coupling in MnSc$_2$S$_4$ decisively influences the magnetic behavior
\cite{giri_2005,kalvius_2006}.
These considerations motivated our investigation using lattice-sensitive
dilatometry and ultrasonic methods, which are also a powerful tool for detecting
magnetic phase transitions.

\begin{figure}[t]
    \includegraphics[width=0.95\columnwidth]{vesta_spinel.png}
    \caption{Crystal structure of MnSc$_2X_4$ with cubic unit cell. Only the
    magnetic Mn sites are shown. These are located so that next-nearest
    neighbors are arranged in triangular planes perpendicular to the [111]
    direction. $J_1$, $J_2$, and $J_3$ represent the magnetic exchange
    interactions between nearest, next-nearest, and third-nearest neighbors,
    respectively. The visualization was done using \textsc{VESTA} \cite{vesta3}.}
    \label{fig:vesta}
\end{figure}

MnSc$_2$Se$_4$ is less extensively investigated, but is related to MnSc$_2$S$_4$,
since the chalcogenide Se$^{2-}$ is only expanding the distances of the magnetic
ions, without changing their magnetic moment or the lattice symmetry \cite{pawlak93}.
Previous results showed similarities to MnSc$_2$S$_4$. Neutron-scattering results
indicated a putative helical ground state with the same propagation vector
{\textit{\textbf{q}}} = (3/4, 3/4, 0) \cite{Guratinder_2022}. In addition,
ac-susceptibility measurements revealed three transitions around 2 K. Our scope
was to map out the magnetic-field-temperature phase diagram of this compound,
compare the results of both compounds, and better understand the link between
possible skyrmions, frustration, and magnetoelastic couplings.

\section{Experimental}
Single crystals of MnSc$_2X_4$ ($X$ = S, Se) with face-centered cubic
structure (space group $Fd\overline{3}m$) were grown by chemical transport reaction, using iodine as the transport agent.
We confirmed the formation of the normal cubic spinel structure by powder
x-ray diffraction on crushed single crystals. 
Both crystals, MnSc$_2$S$_4$ and MnSc$_2$Se$_4$, were of the same purity and quality, as indicated by the powder x-ray diffraction patterns already published in \cite{Fritsch04} and the supplement to \cite{sourd24}. 
We performed magnetic
characterization by SQUID magnetometry and compared the results to that
of stoichiometric powder samples. A more detailed description of the crystal
growth and characterization can be found in Refs. \cite{gao_2017, sourd24}.
We cut pieces with the required orientation optimized for ultrasound as well
as for dilatometry measurements with a characteristic length of 1 to 3 mm
from large crystals. The dimensions of the MnSc$_2$S$_4$ crystal were 2.2 x
2.2 x 1.2 mm$^3$ and 1.65 x 1.65 x 0.85\,mm$^3$ for the MnSc$_2$Se$_4$ crystal.

We performed measurements of the magnetostriction and thermal expansion
parallel to the crystallographic [111] direction with magnetic fields
oriented in the same direction. We used a commercial capacitive dilatometer
cell \cite{kuechler_2012} placed in a $^3$He cryostat equipped with an 18/20 T magnet.
We chose field-sweep rates between 0.05 and 0.25 T/min. The resolution in
relative length changes, $\Delta L/L$, was about 10$^{-7}$.

In addition, we performed ultrasound experiments in the same cryostat,
using a transmission pulse-echo technique with phase-sensitive detection
\cite{Luethi05, Hauspurg24}. We attached LiNbO$_3$ transducers (36°-Y cut
for exciting longitudinal modes) to the polished surfaces of the single
crystals. We applied magnetic fields up to 18 T so that the ultrasound
propagation direction $\textit{\textbf{k}}$ was aligned parallel to the
magnetic field $\textit{\textbf{k}} \parallel \textit{\textbf{H}} \parallel [111]$.

Further, we performed ac-susceptibility measurements down to 1.6 K using
a standard induction setup in a $^{4}$He flow cryostat. We used frequencies
from 500 Hz to 3 kHz and temperature sweep rates of 0.1 to 0.5 K/min.

We determined the specific heat using the quasi-adiabatic heat-pulse method
\cite{Barron1999HeatCA}. For this,
we mounted the sample on a sapphire platform with a calibrated Cernox
thermometer in a $^3$He sorb-pumped cryostat. We measured the heat capacity
of the platform and the grease separately and subtracted
it to determine the specific heat of the sample.

\section{Experimental Results}

\subsection{\texorpdfstring{MnSc$_2$S$_4$}{MnSc2S4}}
\begin{figure}[t]
\centering
  \includegraphics[width=0.48\columnwidth]{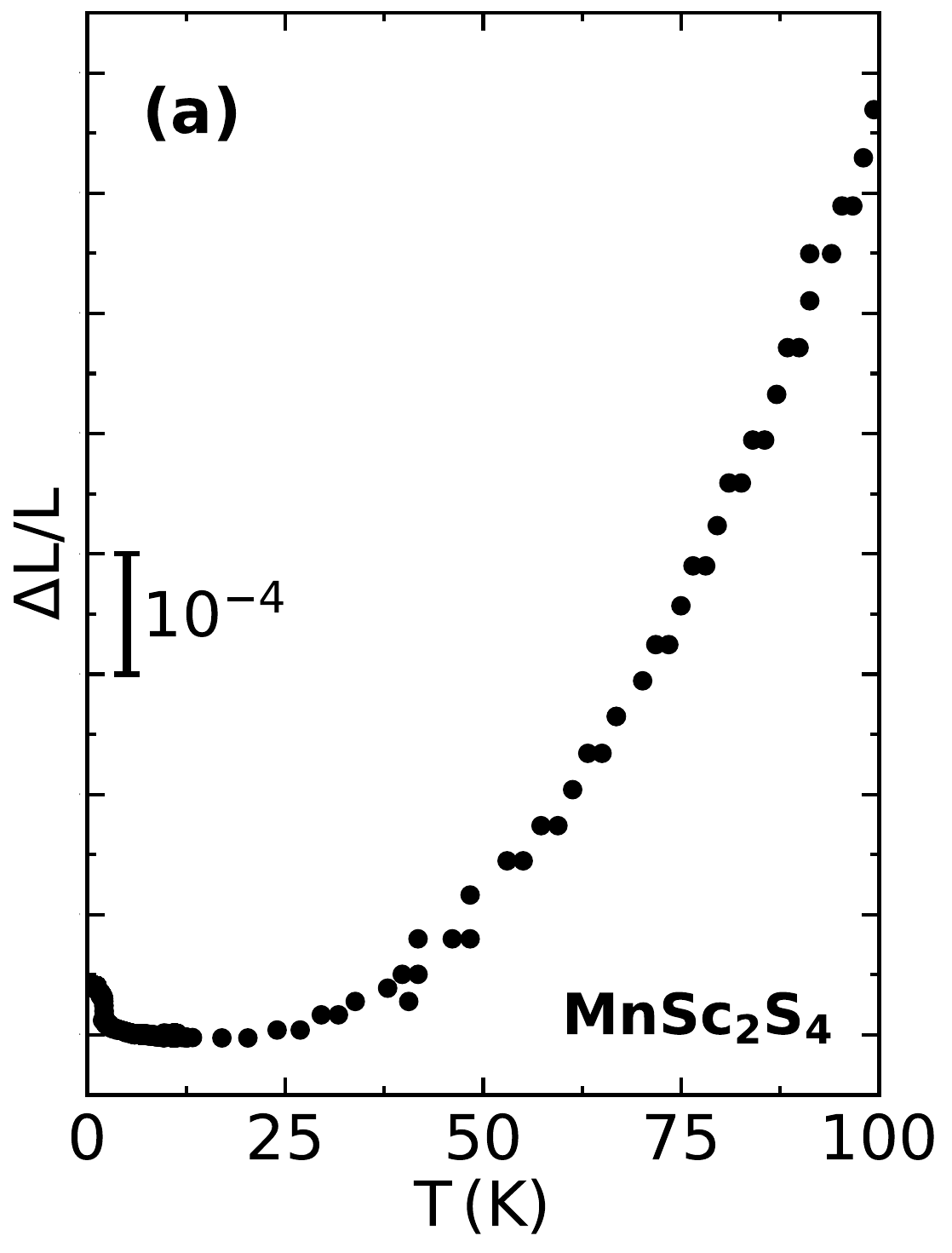}
  \includegraphics[width=0.48\columnwidth]{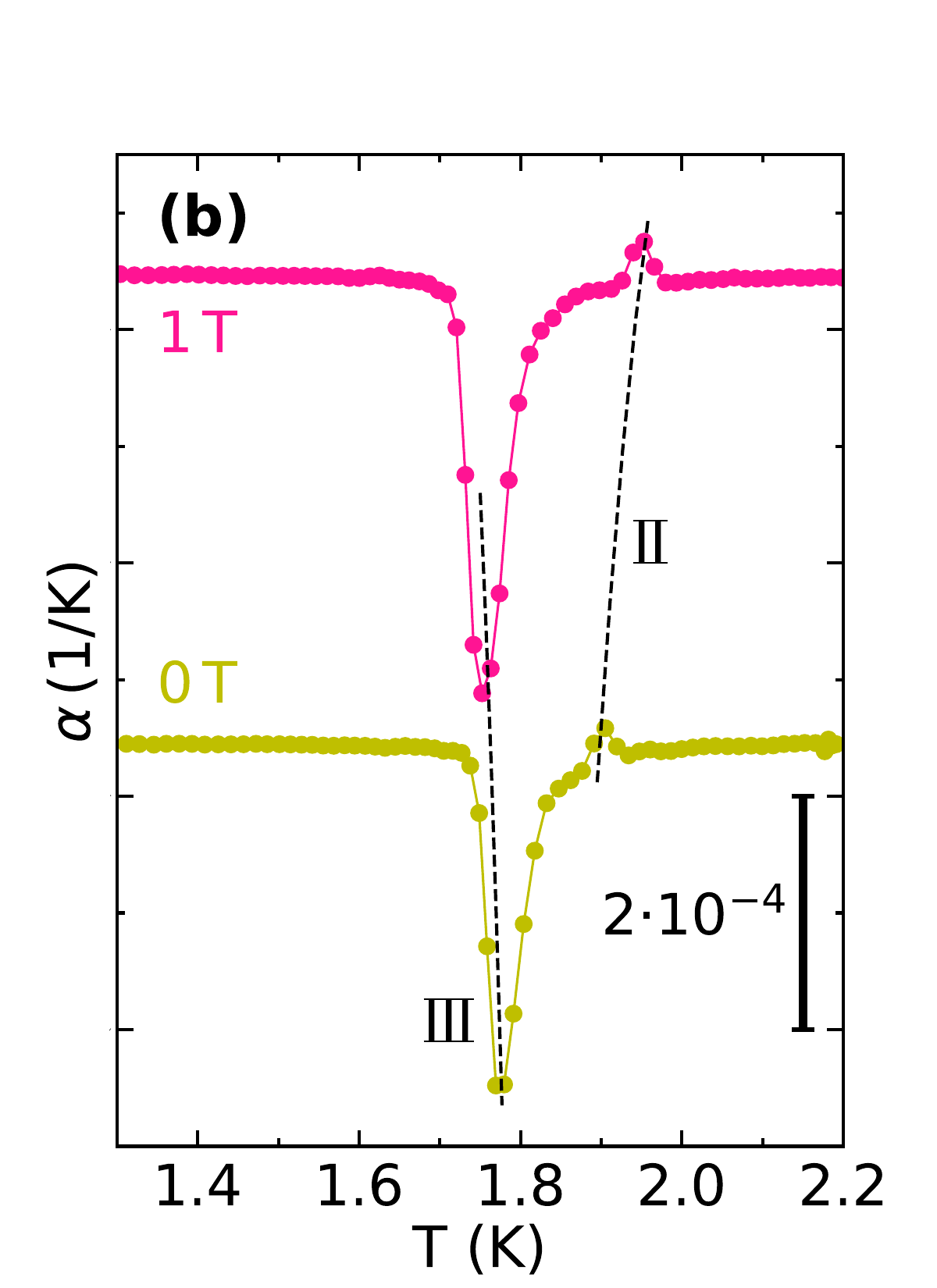}
\caption{(a) Thermal-expansion data of MnSc$_2$S$_4$ up to
100 K. (b) Low-temperature thermal-expansion coefficient $\alpha$ in 0
and 1 T.}
\label{anh}
\end{figure}

\begin{figure}[t]
\centering
\minipage{0.5\columnwidth}
  \includegraphics[width=\linewidth]{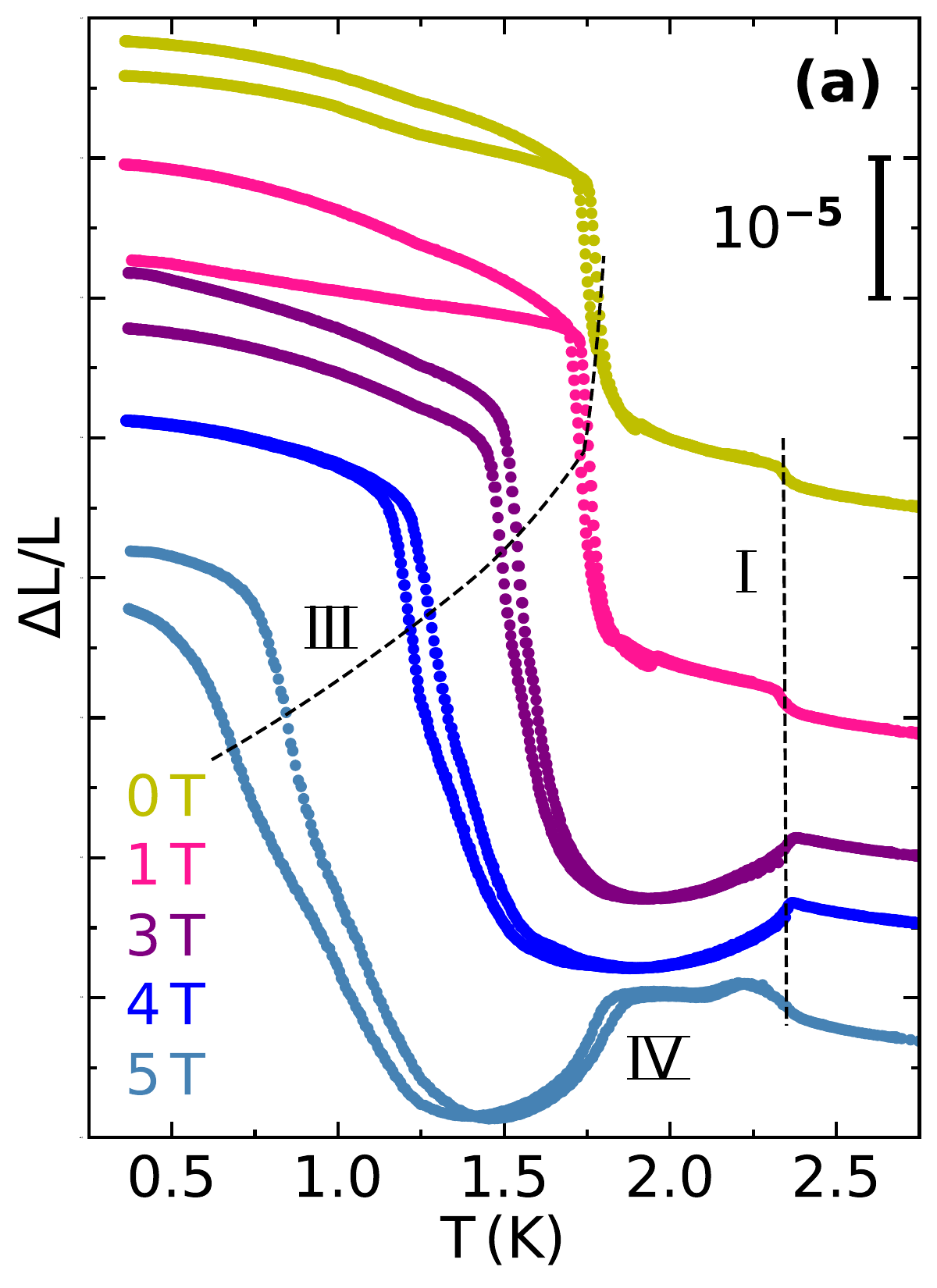}
\endminipage\hfill
\minipage{0.5\columnwidth}
  \includegraphics[width=\linewidth]{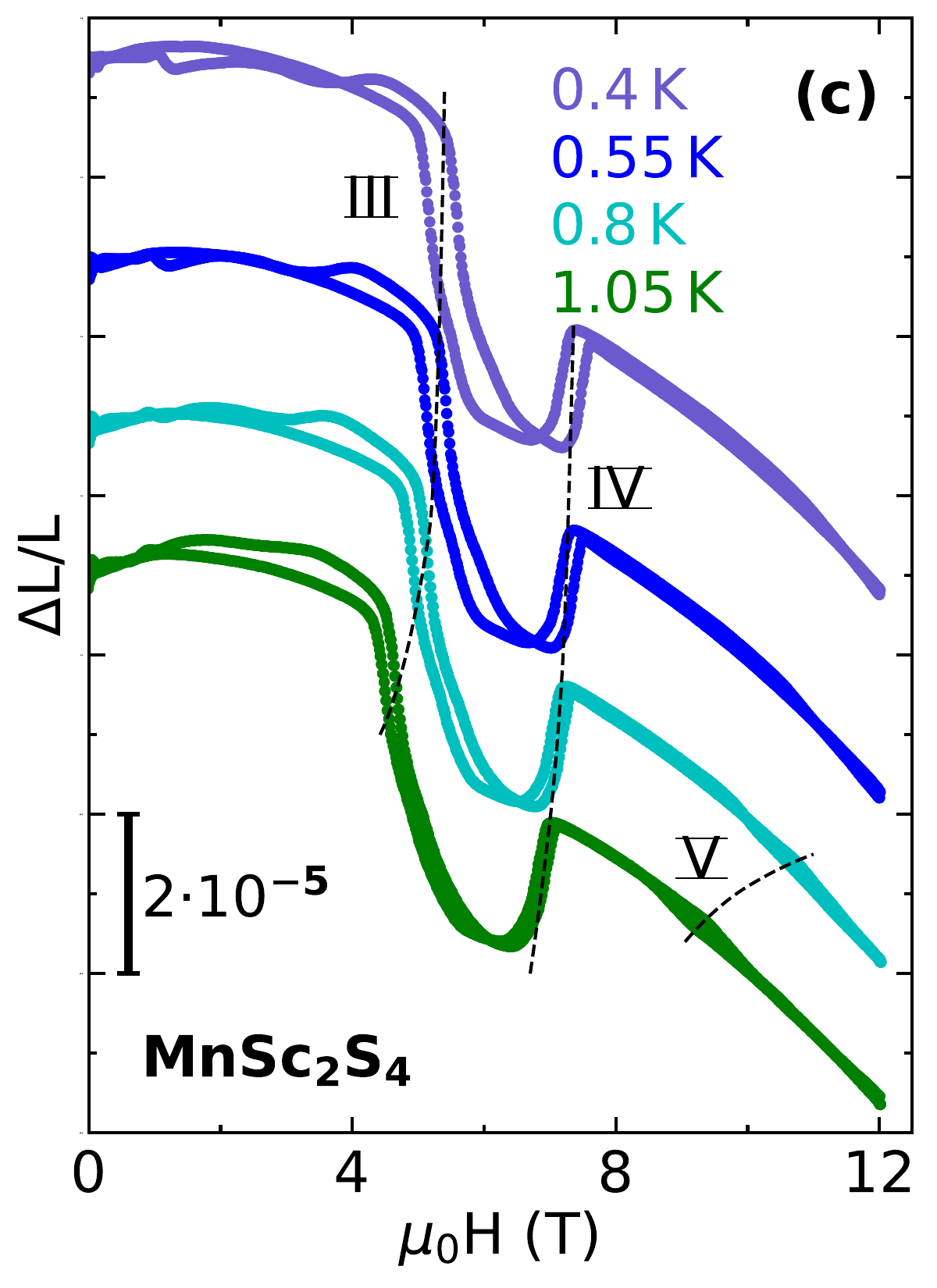}
\endminipage\hfill
\minipage{0.5\columnwidth}
  \includegraphics[width=\linewidth]{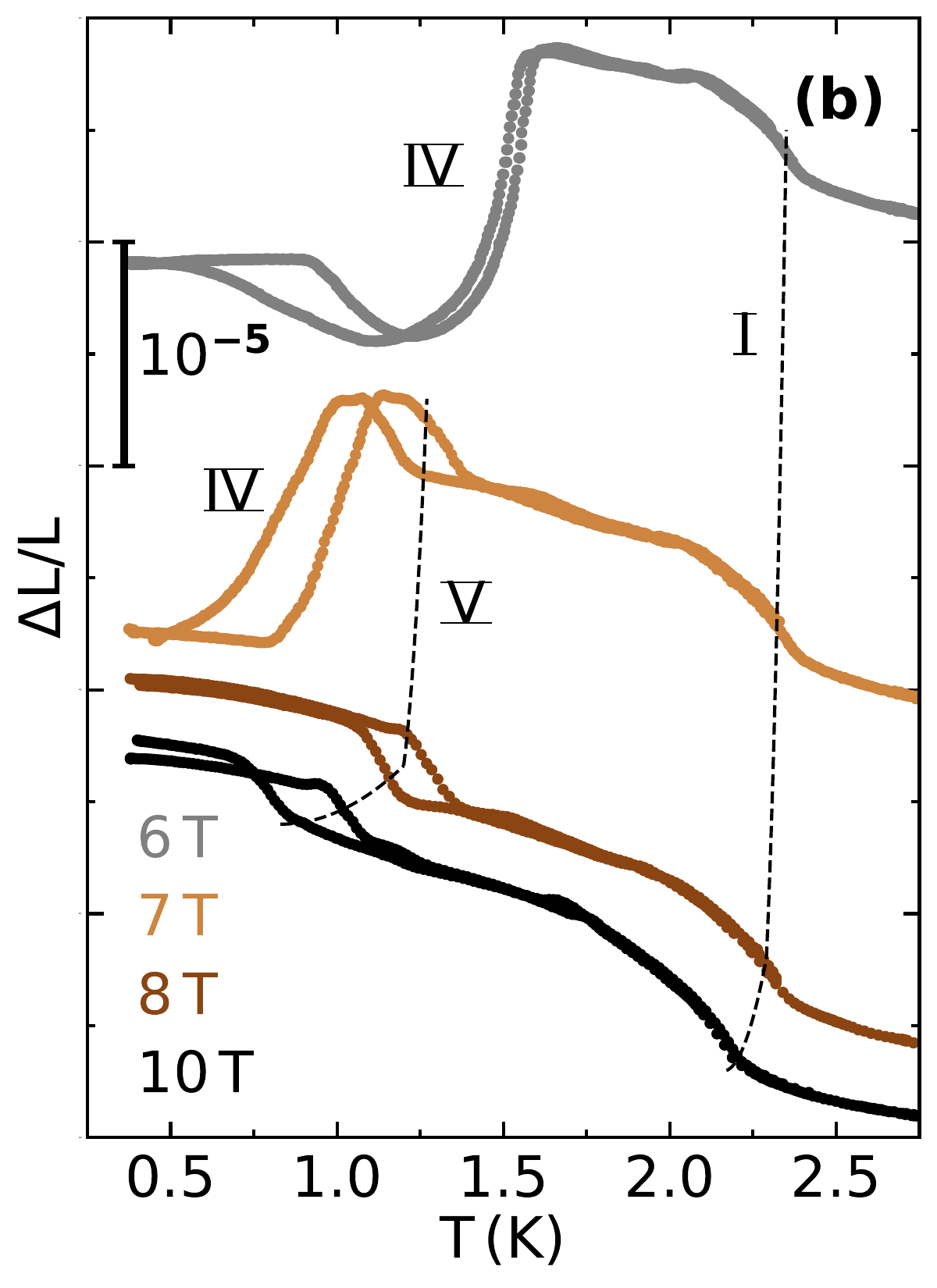}
\endminipage\hfill
\minipage{0.5\columnwidth}
  \includegraphics[width=\linewidth]{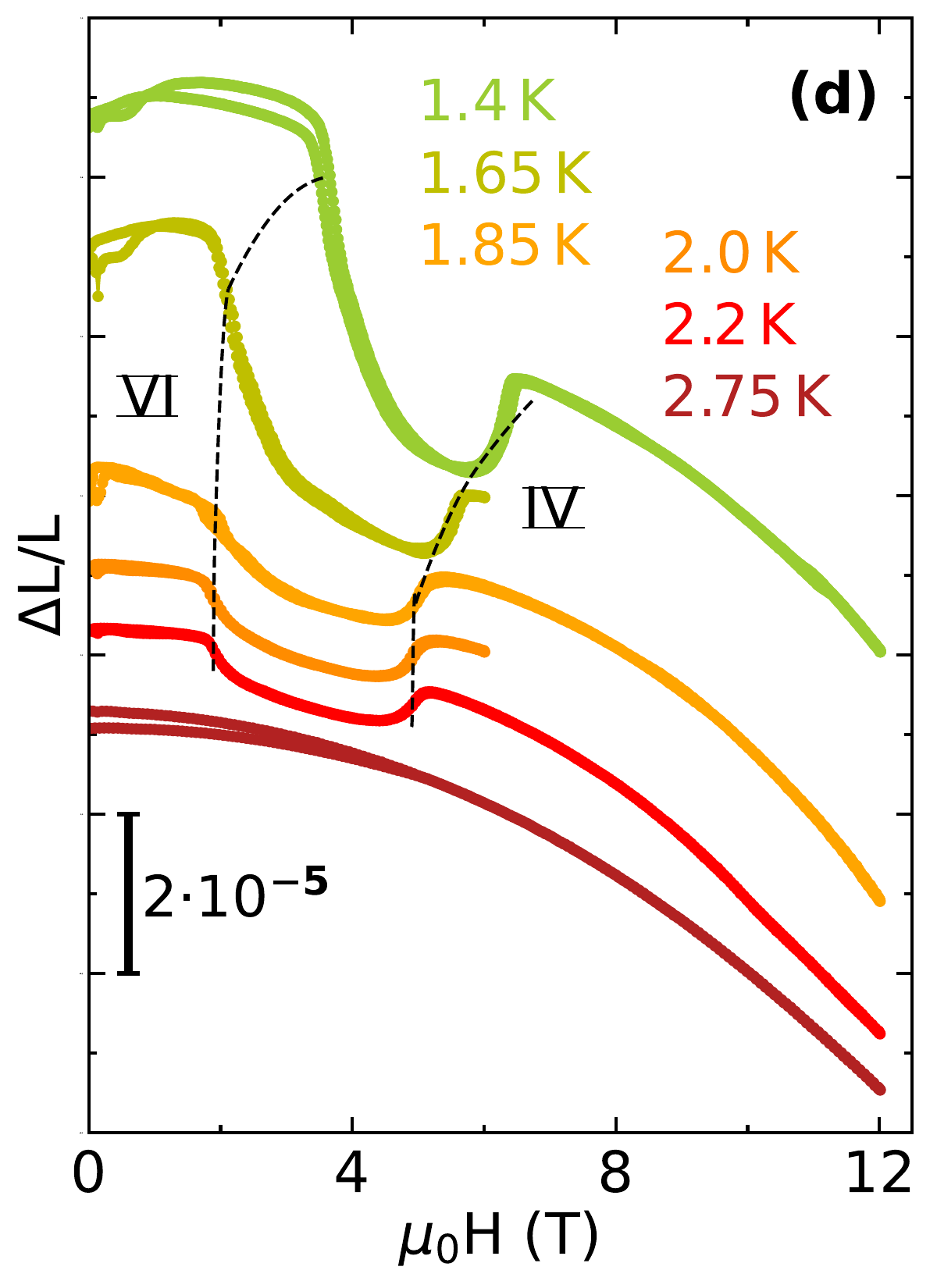}
\endminipage
\caption{(a), (b) Thermal expansion and (c), (d) magnetostriction of
MnSc$_2$S$_4$ single crystals for $H\parallel \Delta L/L \parallel$[111]
after zero-field cooling (ZFC) from the paramagnetic state. The curves are
shifted for better visibility. The black dashed lines indicate the phase
transitions and are guides to the eye.}
  \label{MSS_dil}
\end{figure}

MnSc$_2$S$_4$ shows the usual thermal expansion at high temperatures. In
Fig.\ \ref{anh}(a), we show the relative length change, $\Delta L/L$, between 0.35 and 100 K. Below about 15 K magnetic contributions start to dominate
$\Delta L/L$. We investigated the temperature dependence of $\Delta L/L$
at various magnetic fields applied along [111], as well as the
magnetic-field dependence of the sample length up to 12 T at different
temperatures (Fig. \ref{MSS_dil}).

The thermal expansion [Figs.\ \ref{MSS_dil}(a) and \ref{MSS_dil}(b)] shows
some clear anomalies, which we label by roman numbers. Two of them can be
characterized at zero field at about 1.75 and 2.3 K. The latter (\Romanbar{1}) indicates the ordering temperature $T_N$. At 1.75 K (\Romanbar{3}),
we find a huge magnetoelastic response. This transition is strongly field
dependent and is visible up to about 5 T, where the anomaly broadens and becomes
hysteretic. In the thermal-expansion coefficient $\alpha =$ d$(\Delta L/L)/$d$T$ we observe a weak but clear anomaly slightly above the transition \Romanbar{3}. This transition \Romanbar{2} appears at about
1.9 and 1.95 K at zero and 1 T, respectively [Fig.\ \ref{anh}(b)].
Another anomaly appears at 5 T at 1.8 K (\Romanbar{4}) as a length expansion
with increasing temperature. This anomaly has a similar field dependence as
transition \Romanbar{3} and can be followed up to 7 T. An additional hysteretic
distortion (\Romanbar{5}) shows up at 7, 8, and 10~T near 1 K [Fig.\ \ref{MSS_dil}(b)].

All these anomalies indicate phase transitions. Most of them are also present
in magnetostriction data [Figs.\ \ref{MSS_dil}(c) and \ref{MSS_dil}(d)]. The
transitions \Romanbar{3} and \Romanbar{4} show a strong coupling to the
lattice and form a valley-like region between about 5 and 7 T at lowest
temperatures. This is interesting, because skyrmions are predicted to appear
in this field range \cite{gao_2020}, so this feature requires some further
theoretical analysis (see discussion below). The valley remains up to
2.2 K, while it becomes less pronounced and nearly temperature independent
above 1.65 K, shifting to a slightly changed phase above about 2 T (\Romanbar{6}).
The anomaly \Romanbar{5} can also be detected as a narrow hysteresis at
higher fields. Finally, the data at 2.75 K are completely featureless,
indicating paramagnetism.

\begin{figure}[t]
\centering
\minipage{0.5\columnwidth}
  \includegraphics[width=\linewidth]{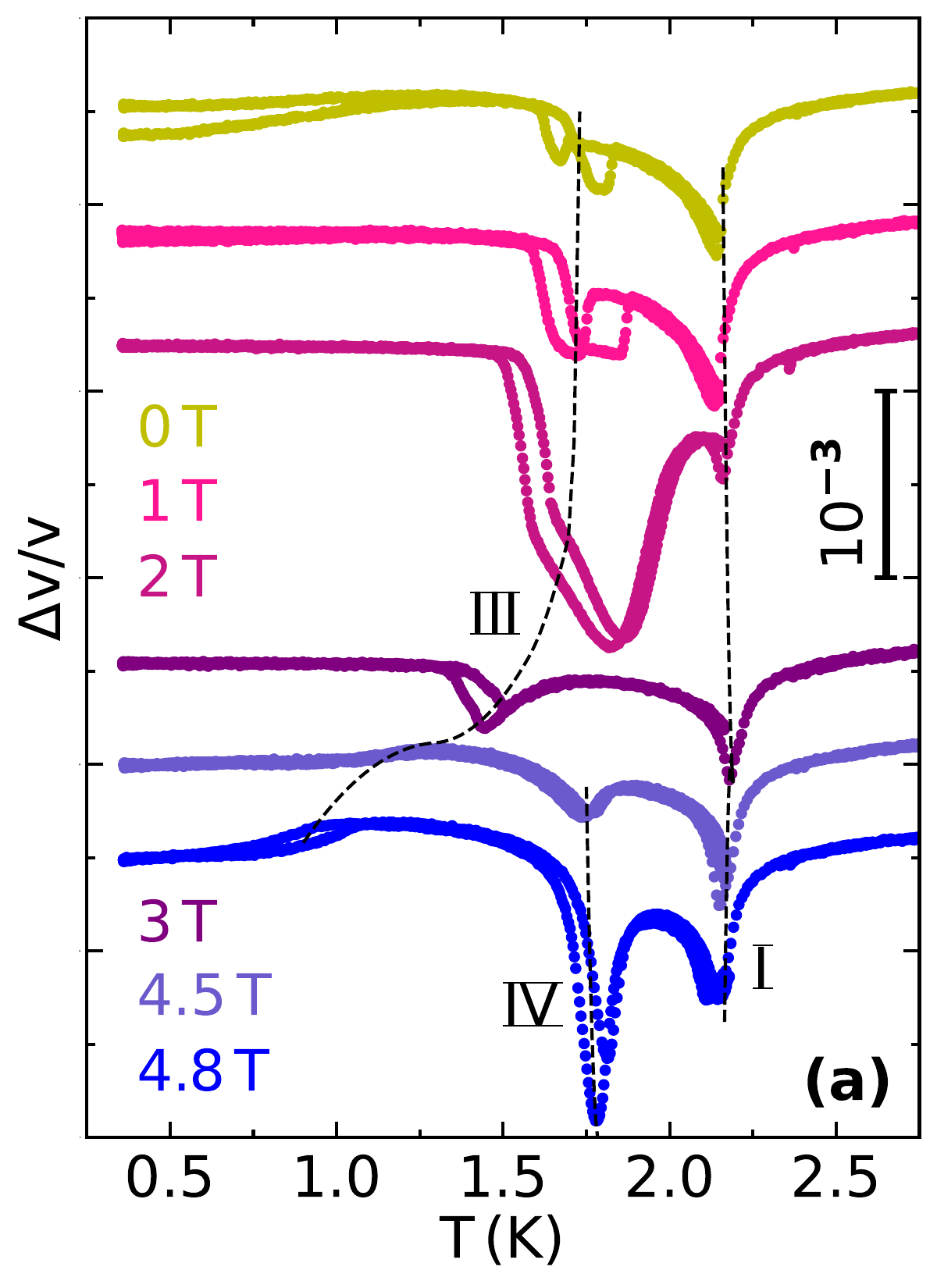}
\endminipage
\minipage{0.5\columnwidth}
  \includegraphics[width=\linewidth]{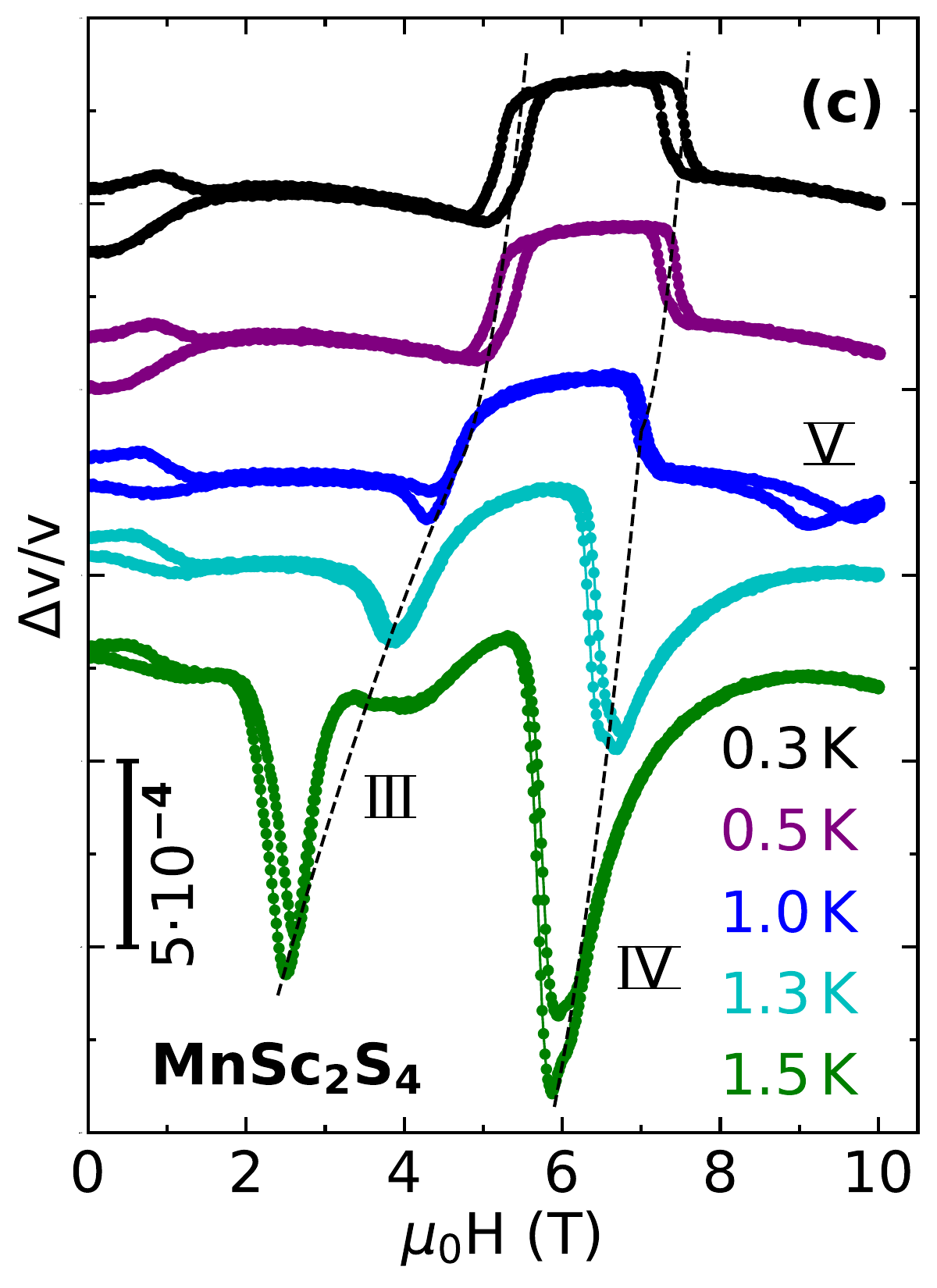}
\endminipage\hfill
\minipage{0.5\columnwidth}
  \includegraphics[width=\linewidth]{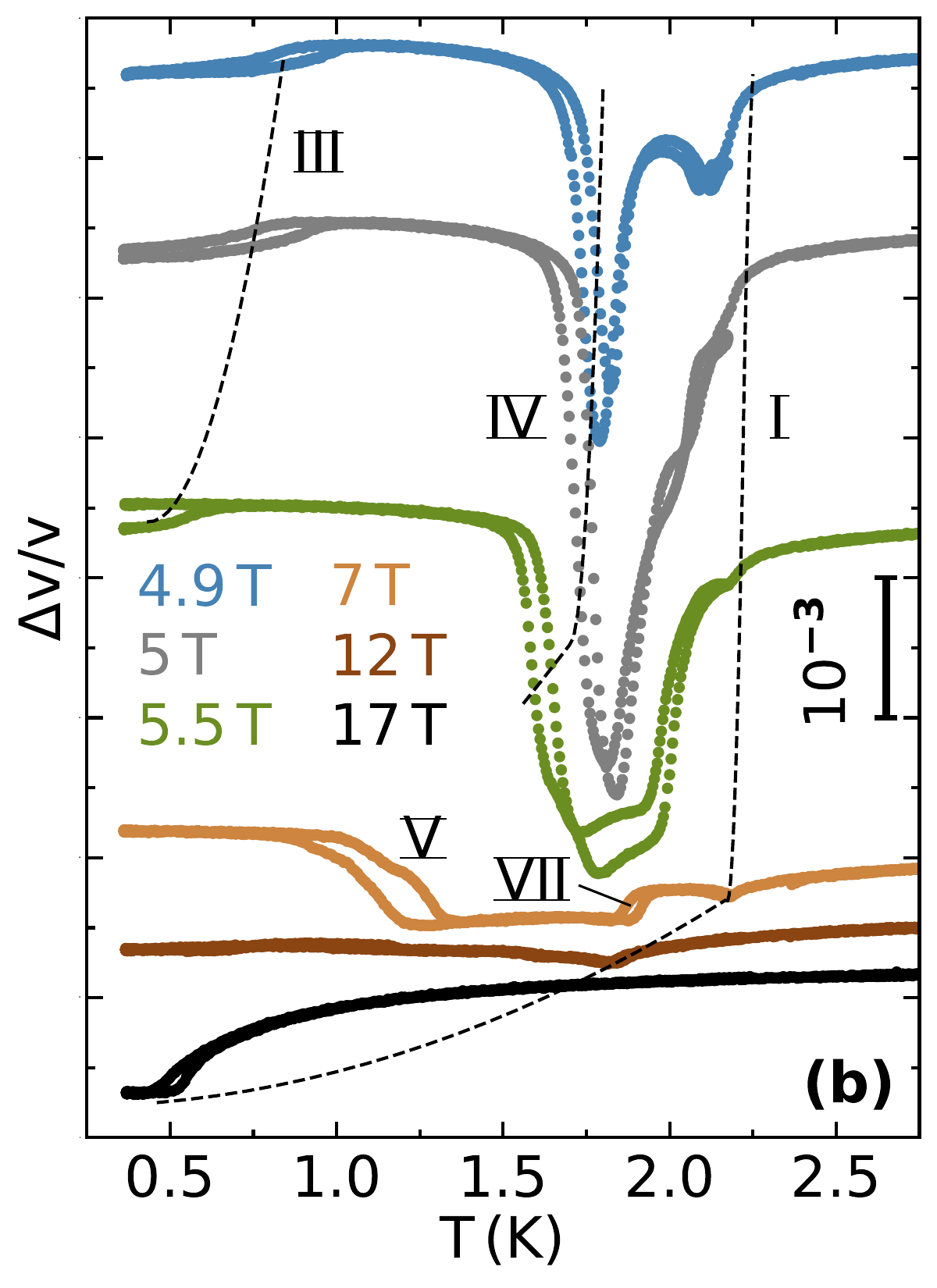}
\endminipage
\minipage{0.5\columnwidth}
  \includegraphics[width=\linewidth]{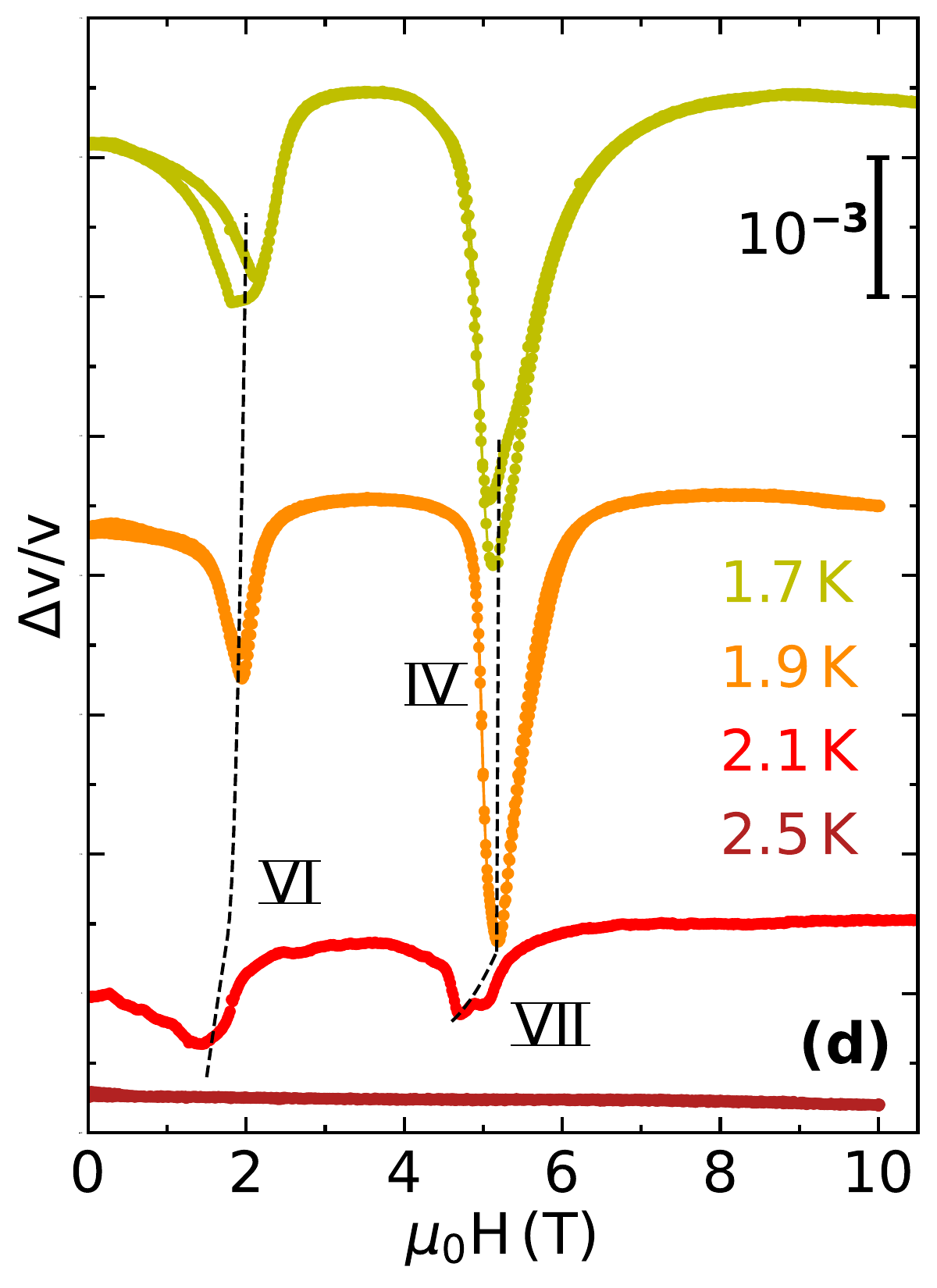}
\endminipage\hfill
\caption{Relative sound velocity of the longitudinal acoustic mode ($\textbf{k}\parallel$\textbf{u}$\parallel\textbf{H}\parallel$[111], f = 81 MHz)
in MnSc$_2$S$_4$ versus (a), (b) temperature and (c), (d) magnetic field,
measured at selected temperatures and magnetic fields, respectively.
The curves are shifted for better visibility. The black dashed lines indicate
the phase transitions and are guides to the eye.}
  \label{fig:US}
\end{figure}

We further performed measurements of the sound-velocity change as a
function of temperature and magnetic field [Fig.\ \ref{fig:US}], analogous
to the dilatometry experiment. We excited a longitudinal mode along [111]
with the magnetic field aligned in the same direction. Figures \ref{fig:US}(a)
and \ref{fig:US}(b) show the relative change of the sound velocity $\Delta v/v$
in MnSc$_2$S$_4$ as a function of temperature for selected magnetic fields.
We observe anomalies at similar temperatures as in our dilatometry
measurements. The transition from the paramagnetic to the magnetically
ordered state (\Romanbar{1}) appears as a clear minimum without any
field dependence up to about 8 T. The transition \Romanbar{3} occurs as
a broad hysteretic feature up to 2 T and changes into a narrow hysteresis
that can be followed up to 5.5~T. We cannot resolve transition \Romanbar{2}
in our ultrasound experiments, which might be caused by the hysteresis of
transition \Romanbar{3}. The transition \Romanbar{4} appears as a distinct
dip that starts at 4.5 T and is visible up to 5.5 T as a sharp jump.
Further, at 7 T, we observe a hysteresis in $\Delta v/v$ at about 1.2 K,
which we attribute to transition \Romanbar{5}. Another anomaly appears at
this field at about 1.9\,K (\Romanbar{7}). The step-like feature
at about 2 K in $\Delta v/v$ at 5.5~T might be related to this transition
as well.

We could observe most of the transitions as well in the field-dependent
ultrasound experiments [Figs.\ \ref{fig:US}(c) and \ref{fig:US}(d)]. The
transitions \Romanbar{3} and \Romanbar{4} dominate the data, enclosing the
previously observed skyrmion phase \cite{gao_2020} as a plateau. Due to their
weak temperature dependence, the transitions \Romanbar{3} and \Romanbar{4}
are sharper in field-dependent ultrasound (and magnetostriction) experiments.
Like in magnetostriction, the shape of the plateau changes at higher
temperatures. Jumps at low magnetic fields continuously turn into minima.
The data at 1 K show a small hysteretic feature around 9 T corresponding
to transition \Romanbar{5}. At 2.1 K, the anomalies \Romanbar{4} and
\Romanbar{7} occur as a double dip at about 5 T, indicating the beginning
of the phase line \Romanbar{7}. Again at 2.5 K, above $T_N$, $\Delta v/v$
is completely featureless in the paramagnetic phase.
Additionally, in Fig. \ref{fig:US}(c) appears a hysteresis below about 1~K for measurements up to 1.5~K. This corresponds to domain orientation.

Both methods compared, it is important to note that the magnitude and shape of some transitions is different for magnetostriction and ultrasound, \textit{e.g.} magnitudes of transitions \Romanbar{4} and \Romanbar{6}. Despite being lattice sensitive, both methods use different concepts to track lattice-effects. This leads to different sensitivities and overall shapes of the anomalies.

\begin{figure}[tb]
\includegraphics[width=0.98\columnwidth]{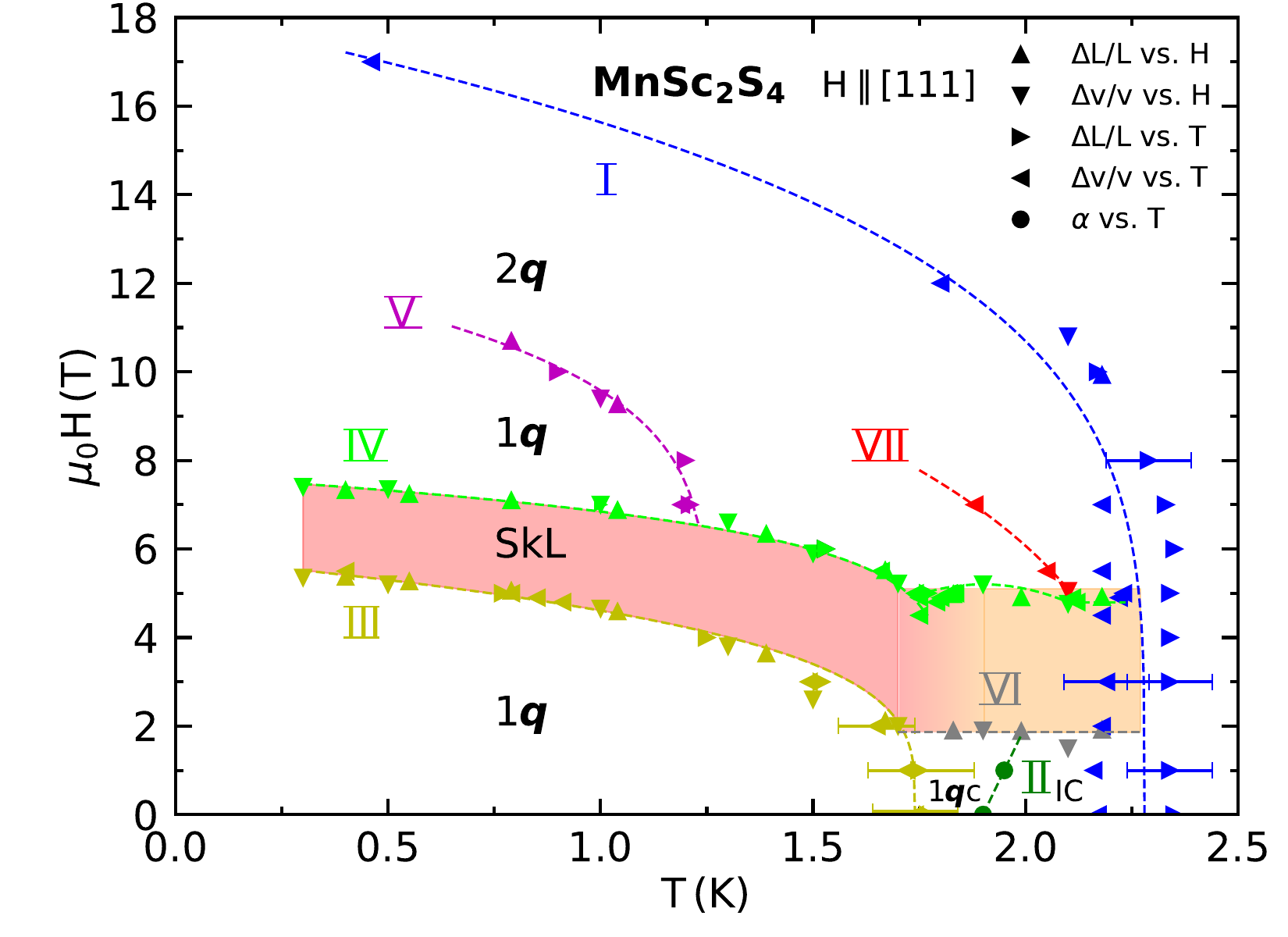}
\caption{Phase diagram of MnSc$_2$S$_4$. The numbers correspond to the
labeling of the anomalies used in the text. The abreviations indicate the magnetic structure of the phases (see also schematic pictures in \cite{gao_2017}). The shape of the markers
indicates the method used to extract the transitions.}
\label{fig:phasediagram}
\end{figure}

From the anomalies observed in the magnetoelastic data, we can construct
the magnetic ($H,T$) phase diagram for fields along [111], shown in
Fig.\ \ref{fig:phasediagram}. Our ultrasound and dilatometry data reveal a
N\'{e}el temperature of 2.3 K. Due to the different measurement configurations, we see a small temperature difference within the experimental error of 5\%. Error bars have been added to some of the affected triangles. The N\'{e}el temperature is ultimately in line with
earlier susceptibility ($T_N = 2.3$~K) \cite{krimmel_2006} and specific-heat
data with the peak at about
2.1 K \cite{Fritsch04}. 
Our temperature-dependent data allow to follow this phase transition \Romanbar{1} towards larger magnetic fields.
Towards lower temperatures, we observe two further
anomalies in the thermal expansion $\alpha$ in zero field [Fig.\ \ref{anh}(b)],
in line with neutron-scattering data \cite{gao_2017, gao_2020}. The first
transition (\Romanbar{2}) is very weak in $\alpha$, whereas transition
\Romanbar{3} leads to a large lattice expansion towards lower temperatures
[Figs.\ \ref{anh}(b) and \ref{MSS_dil}(a)]. The ultrasound data [Fig.\
\ref{fig:US}(a)] show a pronounced hysteresis in this temperature region.
The error bars shown in Fig.\ \ref{fig:phasediagram} reflect the width of
this hysteresis.

In applied magnetic fields, our dilatometry and ultrasound data show
anomalies at the phase transitions \Romanbar{3} and \Romanbar{6}. These
anomalies agree well with the transitions to the 3\textit{\textbf{q}} or
antiferromagnetic skyrmionic phase observed in neutron scattering (red
colored area in Fig.\ \ref{fig:phasediagram}) \cite{gao_2020}.

At higher fields, above the skyrmion phase, two additional phase lines
(\Romanbar{5} and \Romanbar{7}) occur. The first one (\Romanbar{5}),
below 1.2 K may separate a fan-single-\textit{\textbf{q}} structure from
a 2\textit{\textbf{q}} structure. Our mean-field calculations, discussed
later, suggest such a 2\textit{\textbf{q}} phase. The anomaly \Romanbar{7}
appears only in our ultrasound data. For this, we have no theoretical
interpretation. It also seems that this line fades away towards lower
temperatures and higher fields.

\subsection{\texorpdfstring{MnSc$_2$Se$_4$}{MnSc2Se4}}
\begin{figure}[t]
\centering
\minipage{0.5\columnwidth}
  \includegraphics[width=\linewidth]{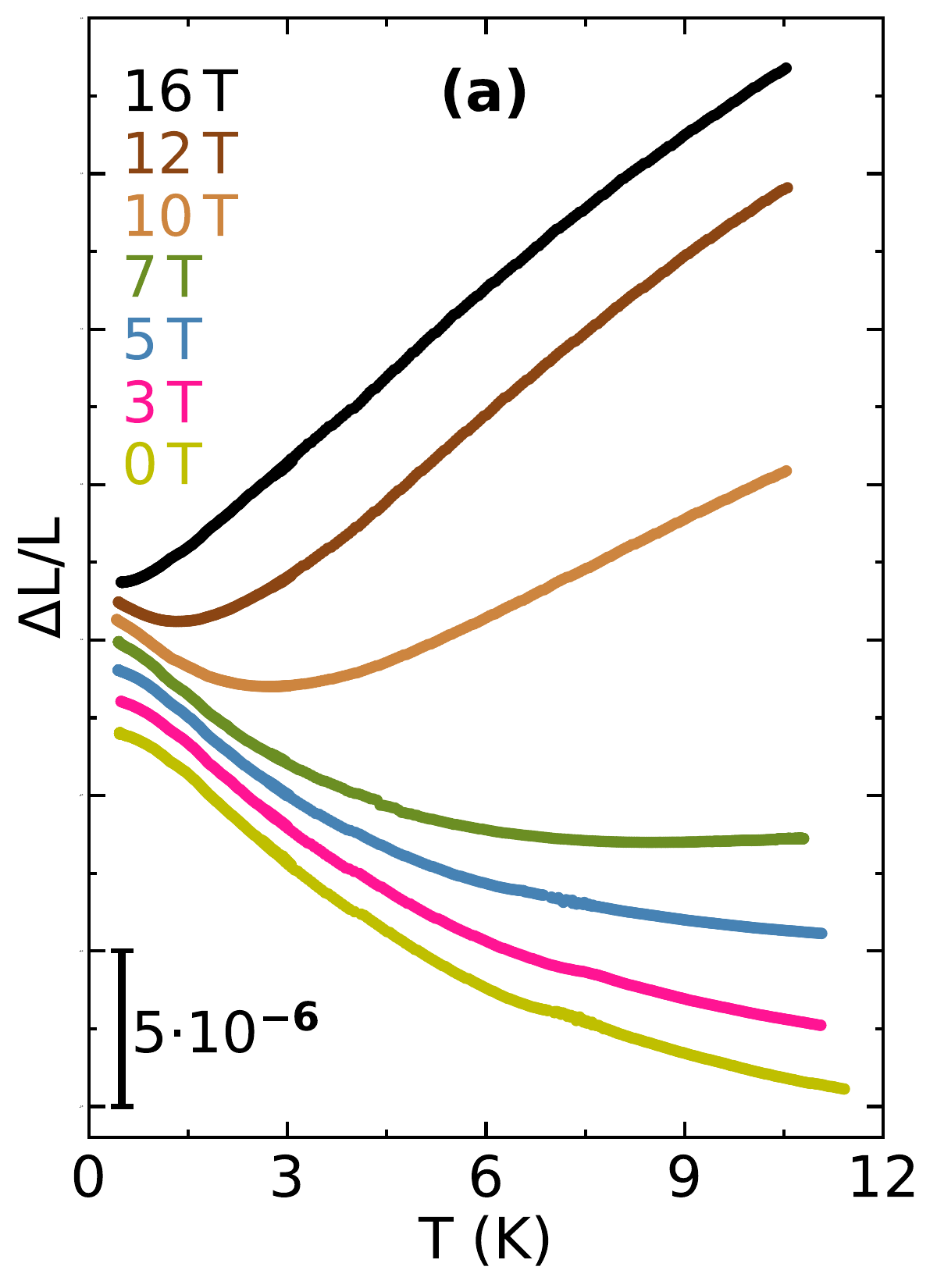}
\endminipage\hfill
\minipage{0.5\columnwidth}
  \includegraphics[width=\linewidth]{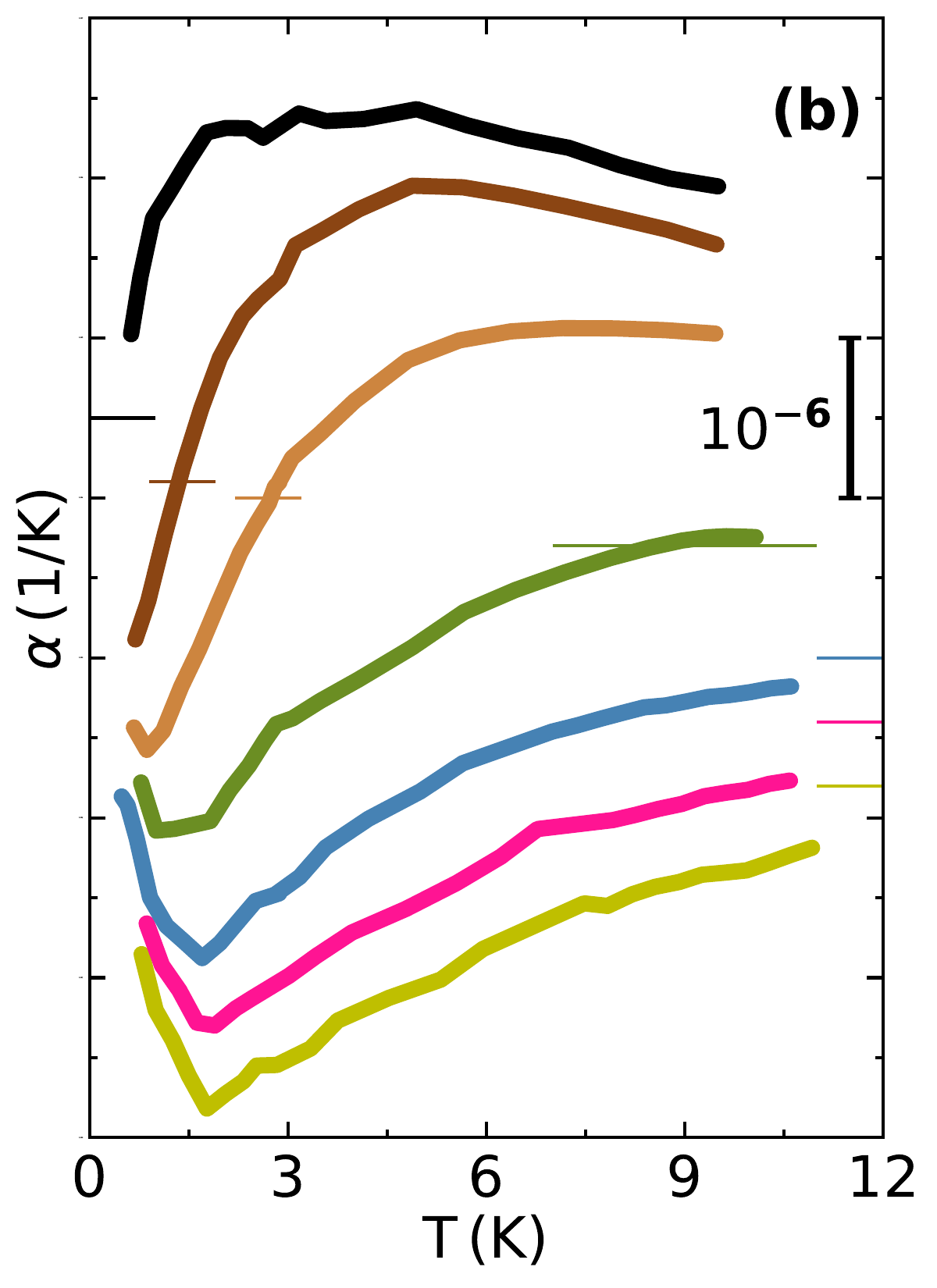}
\endminipage\hfill
\minipage{0.5\columnwidth}
  \includegraphics[width=\linewidth]{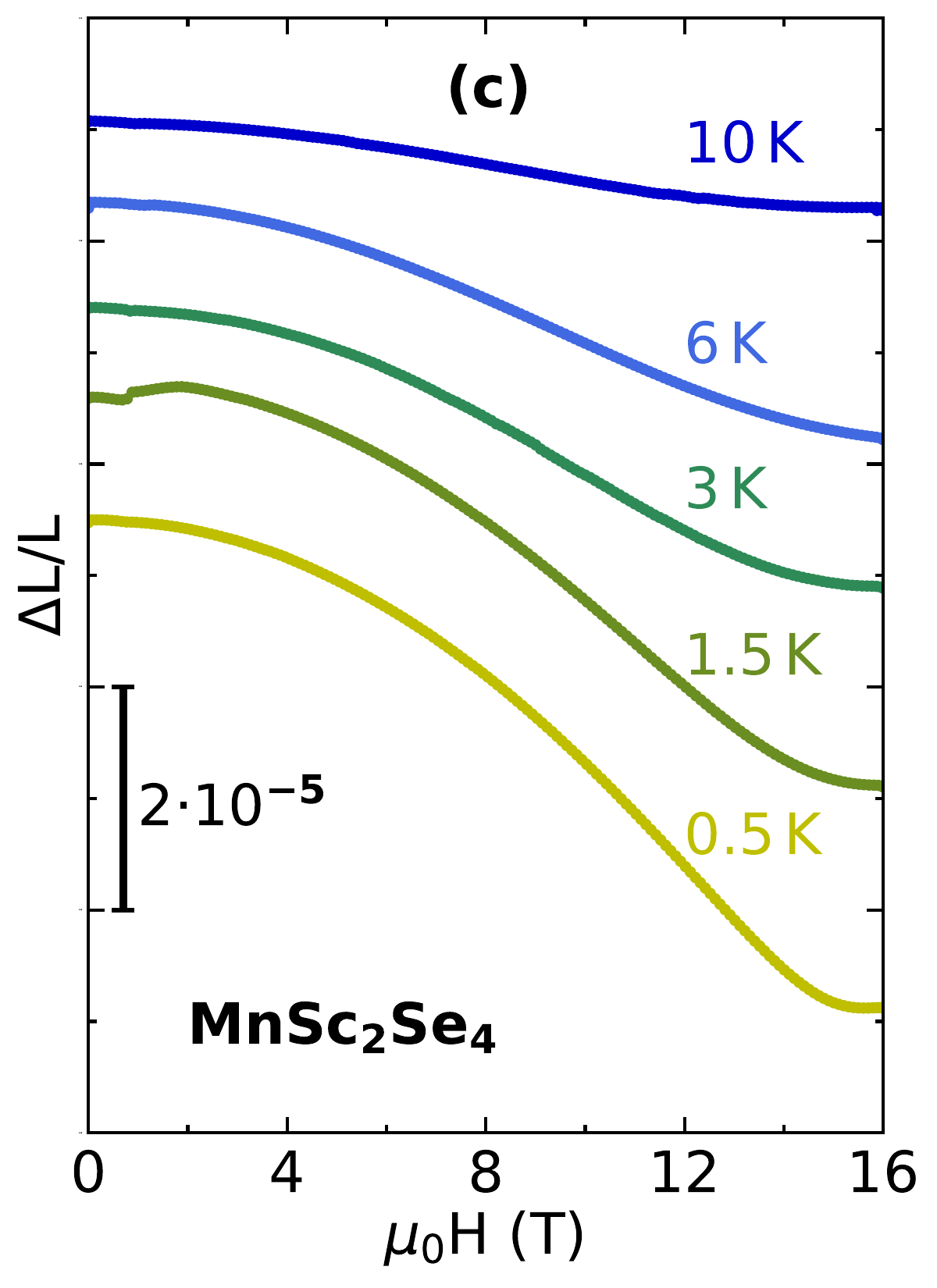}
\endminipage\hfill
\minipage{0.5\columnwidth}
  \includegraphics[width=\linewidth]{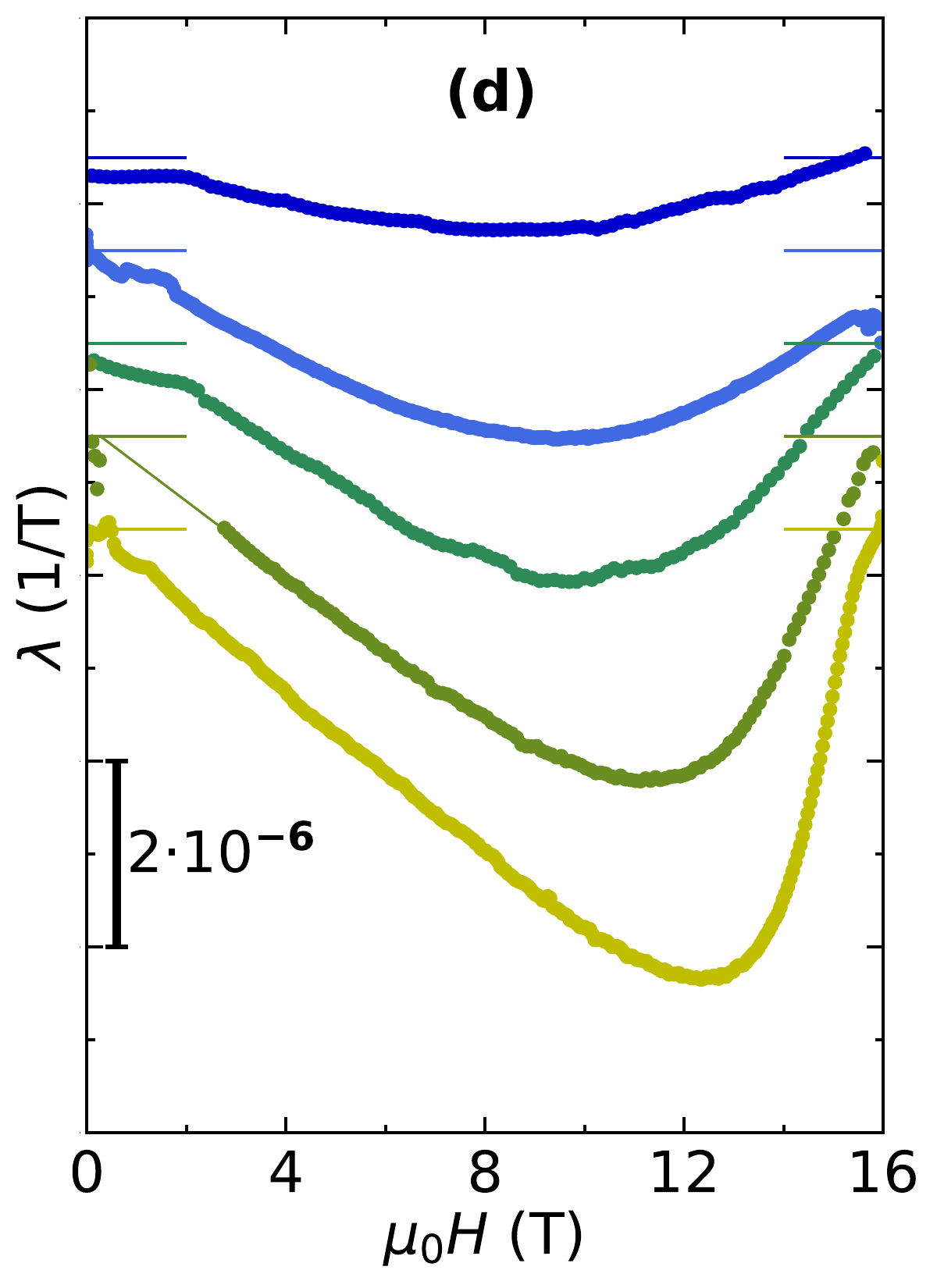}
\endminipage
\caption{Dilatometry data of MnSc$_2$Se$_4$ in longitudinal geometry ($\Delta L/L\parallel$ \textit{\textbf{H}} $\parallel$ [111]). (a) Thermal expansion and
(b) its derivative, the thermal expansion coefficient $\alpha$.
(c) Magnetostriction and (d) its derivative, the magnetostriction coefficient
$\lambda$, up to 16\,T. All curves are shifted for better visibility. The
horizontal lines in the derivatives mark the zero point of the coefficient
for the respective curve.}
  \label{sedil}
\end{figure}

For MnSc$_2$Se$_4$, we performed as well dilatometry and ultrasound
measurements as a function of temperature and magnetic field. 
Note that we used a single crystal of similar quality compared to MnSc$_2$S$_4$.
Under
similar experimental conditions, we observed that MnSc$_2$Se$_4$ exhibits
significantly fewer features than MnSc$_2$S$_4$. The temperature-dependent
dilatometry measurements at various fixed magnetic fields reveal curves
with negative slopes for moderate fields up to 5 T [Fig.\ \ref{sedil}(a)].
At higher fields, we find broad minima that shift to lower temperatures.

The field-dependent length changes [Fig.\ \ref{sedil}(c)] show negative
slopes as well. As the temperature decreases, the magnetostrictive effect
becomes stronger, indicating a quadratic parastriction that dominates over
the thermal lattice vibrations. A model, based on exchange striction as the
dominant interaction, can explain the negative slope in both experiments
\cite{doerr_magnetostriction_2005}.

In order to check for some minor, hidden anomalies, we calculated the
derivatives $\alpha $ [Fig.\ \ref{sedil}(b)]
and $\lambda$~=~$\frac{1}{\mu_0}$d$(\Delta L/L)$/dH [Fig.\ \ref{sedil}(d)].
However, also the thermal-expansion coefficient $\alpha$ is mostly featureless.
For smaller fields, a broad minimum appears up to about 10 T. This minimum
is temperature dependent, shifting from about 1.7 K at zero field to about
0.9 K at 10 T and is suppressed at 12 T. In the magnetostriction coefficient
$\lambda$, a broad minimum appears as well, which shifts to higher magnetic
fields with decreasing temperature. Ultimately, the dilatometry data show no clear
anomaly that would indicate a phase transition.

\begin{figure}[t]
\centering
\minipage{0.5\columnwidth}
  \includegraphics[width=\linewidth]{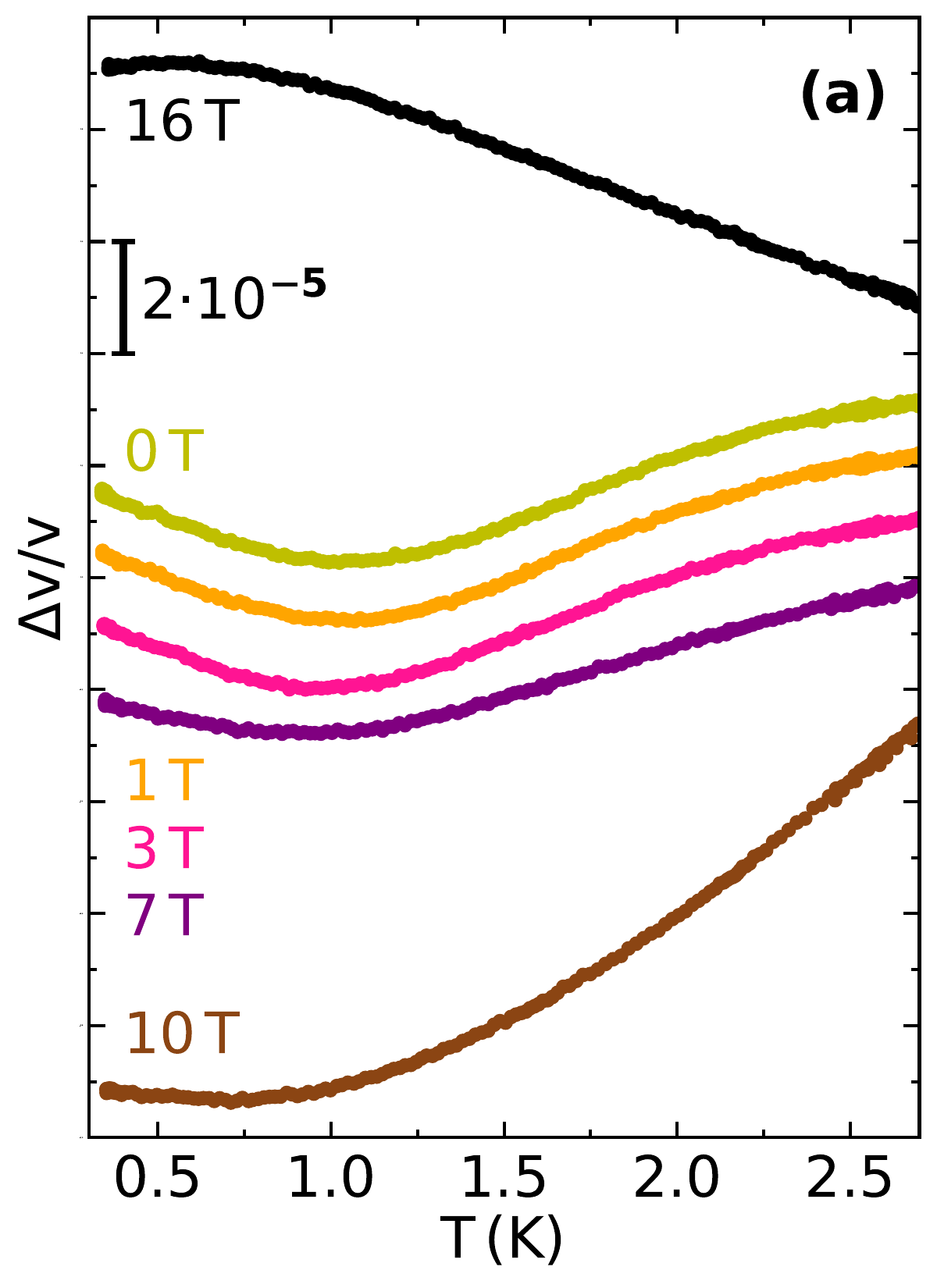}
\endminipage\hfill
\minipage{0.5\columnwidth}
  \includegraphics[width=\linewidth]{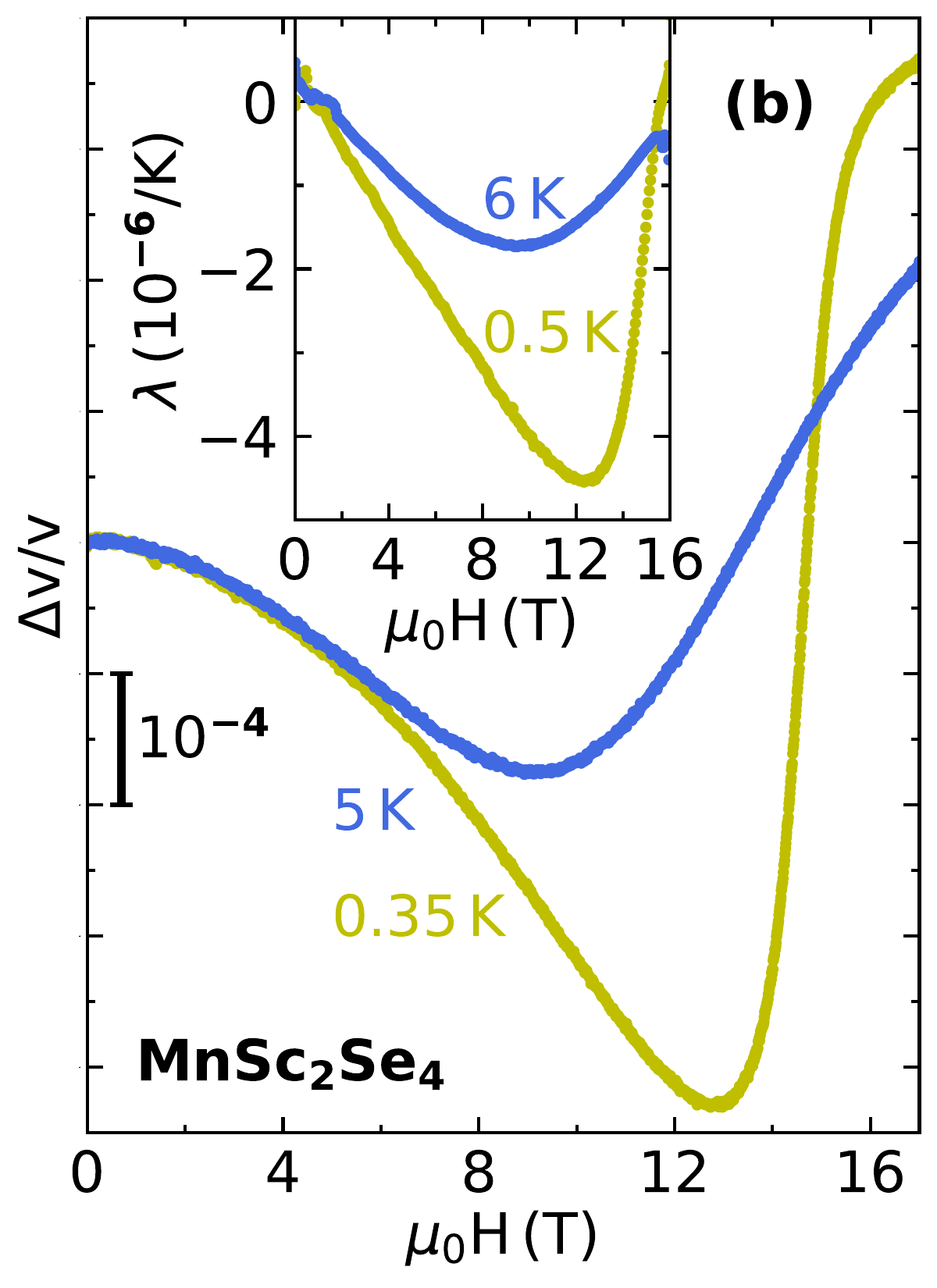}
\endminipage
\caption{(a) Temperature-dependent and (b) field-dependent sound-velocity
change in MnSc$_2$Se$_4$ for the longitudinal acoustic mode (\textit{\textbf{k}}
$\parallel$ \textit{\textbf{u}} $\parallel$ \textit{\textbf{H}} $\parallel$ [111],
$f=81$ MHz) at selected magnetic fields \cite{sourd24}. The inset in (b) shows
the magnetostriction coefficient $\lambda = \partial/\partial B (\Delta L/L)$,
recorded at similar temperatures [see also Fig.\ \ref{sedil}(d)].}
  \label{usse}
\end{figure}

We also measured the sound velocity changes as a function of $T$ and
$H\parallel [111]$ for the longitudinal mode \textit{\textbf{k}}
$\parallel$ \textit{\textbf{u}} $\parallel$ [111].
Figure \ref{usse}(a) shows the relative change of the sound-velocity
$\Delta v/v$ in MnSc$_2$Se$_4$ as a function of temperature for selected
magnetic fields. For MnSc$_2$S$_4$, we found anomalies in $\Delta v/v$ in
the order of 10$^{-3}$. Here, we could not observe any sharp feature down
to a resolution of some 10$^{-6}$. In the temperature-dependent measurements,
we observe a phonon softening down to about 1~K, which initially becomes larger
with increasing magnetic field. At highest field (16 T), this dependence is
reversed with a temperature-dependent hardening and saturation of $\Delta v/v$
below about 0.7 K.

These results are in accordance with the sound-velocity data measured as a
function of magnetic field [Fig.\ \ref{usse}(b)]. At 0.35 K, we find a
broad minimum at about 13 T. this minimum broadens and shifts to lower
fields at higher temperatures. Sourd \textit{et al.} have discussed this behavior
in detail previously \cite{sourd24}. When comparing these data with the
magnetostriction coefficient $\lambda$ in the inset of Fig. \ref{usse}(b)
[complete set of curves in Fig.\ \ref{sedil}(d)] the similar behavior of
the broad anomalies is striking.

\begin{figure}[t]
\includegraphics[width=0.98\columnwidth]{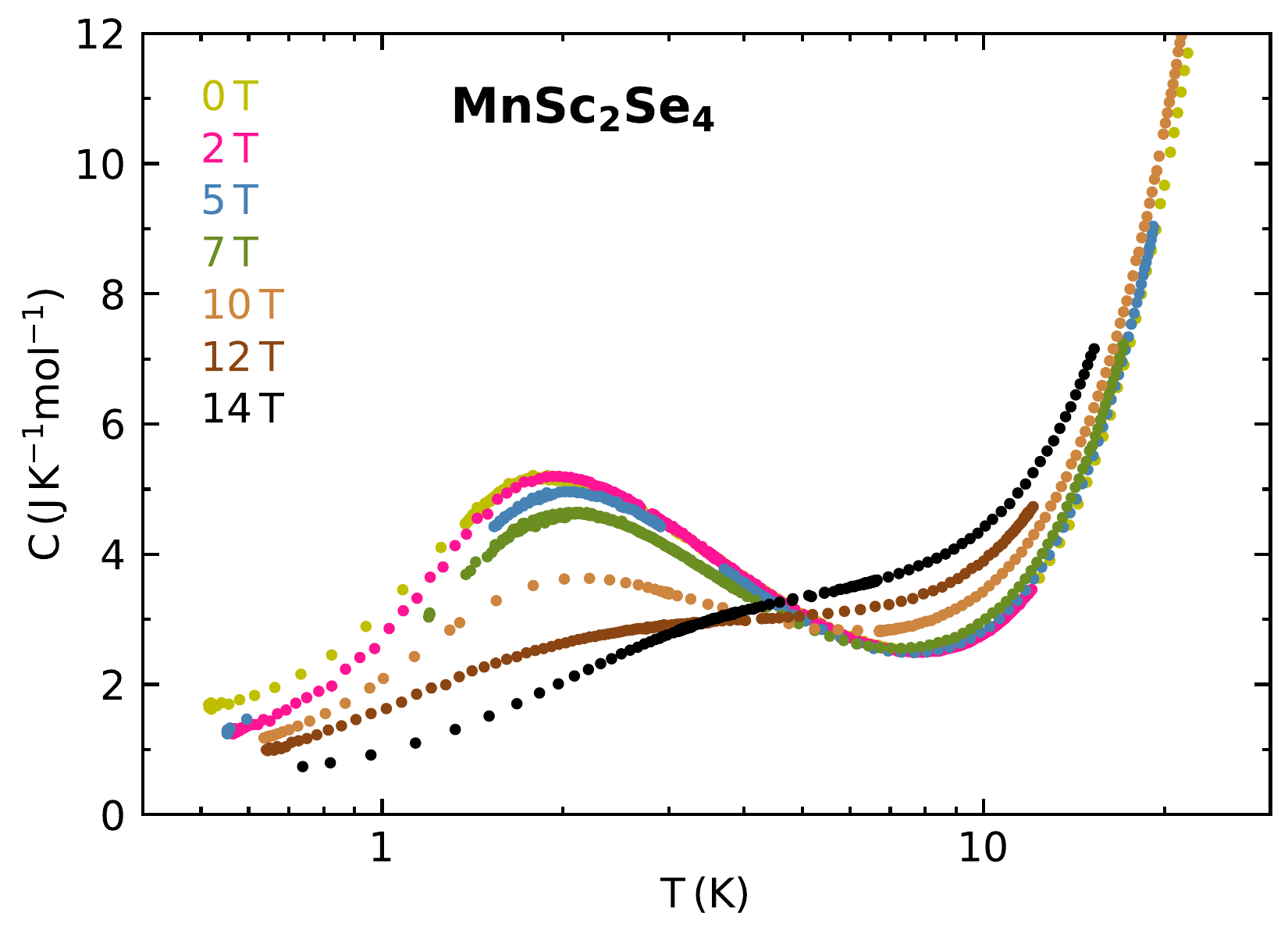}
\caption{Specific heat of MnSc$_2$Se$_4$ in various fields applied along the
[111] direction.}
\label{spech}
\end{figure}

In order to search further for possible phase transitions in MnSc$_2$Se$_4$,
we performed some additional thermodynamic investigations.
We measured the specific heat in magnetic fields up to 14 T (Fig. \ref{spech}).
In zero field, the specific heat shows a broad anomaly at about 2 K. With
increasing magnetic field, this feature becomes smaller, until it vanishes at
about 12 T with the magnetic entropy shifting to higher temperatures.
The specific-heat data evidence the absence of a magnetic phase transition
down to 0.5 K.

\begin{figure}[b]
\includegraphics[width=0.98\columnwidth]{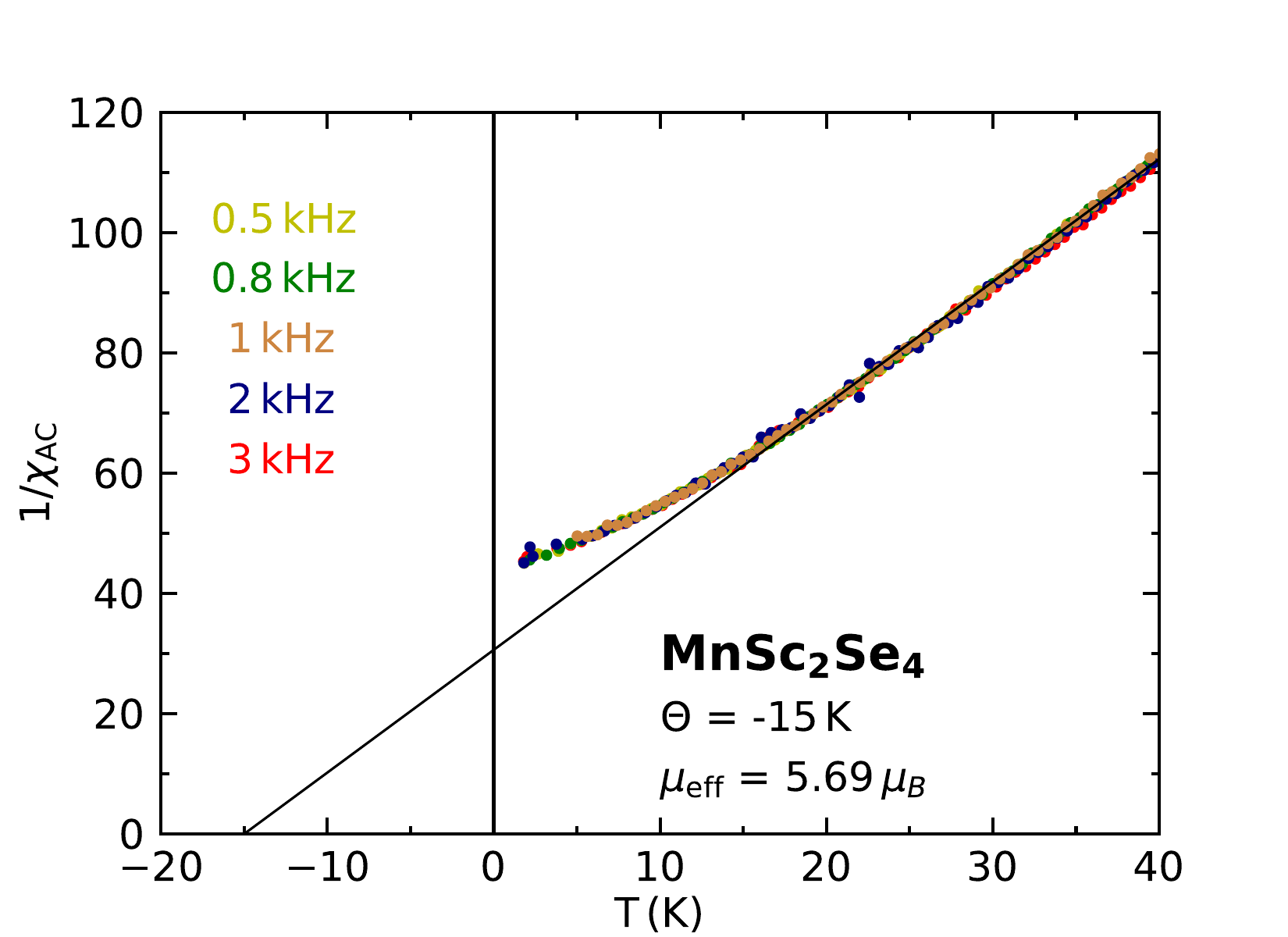}
\caption{Temperature-dependent inverse ac susceptibility of MnSc$_2$Se$_4$
in zero field measured using various excitation frequencies of the ac field
applied parallel to the [111] direction.
The black line shows a linear fit to the data above 20 K.}
\label{sus}
\end{figure}

We further measured the ac susceptibility in zero field at various excitation
frequencies down to 1.5 K. The inverse ac susceptibility follows nicely a
Curie-Weiss behavior at high temperatures, but deviates from the linear curve
below about 20 K (Fig. \ref{sus}). Such a deviation is characteristic for the
existence of short-range (antiferromagnetic) correlations. The paramagnetic
Curie-Weiss temperature is $\Theta_{CW}$ = -15 K and the effective moment is
$\mu_{eff}$ = 5.69 $\mu_B$. These values are in line with previously reported
values of $\Theta_{CW}$ = -18.4 K and $\mu_{eff}$ = 5.81 $\mu_B$ extracted
from ac-susceptibility data measured up to 400~K on a polycrystalline sample \cite{Guratinder_2022}.
$\mu_{eff}$ is close to 5.92 $\mu_B$ for pure Mn$^{2+}$ $S$ = 5/2 spins.

The data show no frequency dependence between 0.5 and 3 kHz, which rules out spin-glass
behavior. Our measurements would also agree with the existence of a
spin-liquid state.

Comparing the results for both compounds, we find a largely different magnetic
behavior. On the one hand, for MnSc$_2$S$_4$, we find a rich phase diagram
including an antiferromagnetic skyrmion state above 2 to 5 T below about
2.3 K. In contrast, MnSc$_2$Se$_4$ shows no indication of a phase transition
down to 0.3 K and up to 16 T. The substitution of the smaller S$^{2-}$ ions
by the larger Se$^{2-}$ ions seems to weaken the interatomic coupling between
the magnetic Mn ions, which prevents the formation of long-range order.

\section{Discussion}
\subsection{\texorpdfstring{MnSc$_2$S$_4$}{MnSc2X4}}
In addition to experimental investigations, we used the mean-field
simulation program \textsc{McPhase} \cite{mcphasemanual} to rationalize the strong magnetoelastic
response to skyrmions. We extended the theory introduced
by Gao \textit{et al.}\ \cite{gao_2020} by an elastic-energy term to calculate the
lattice expansion. We set up a model taking into account only effects of
exchange striction, since we expect only very weak second-order crystal-field
effects for the Mn$^{3+}$ ions. Based on our notation the Hamiltonian appears as
\begin{equation}
\begin{split}
  \mathcal{H} 
    &= \mathcal{H}_0 + \mathcal{H}_{\parallel} + \mathcal{H}_A + \mathcal{H}_Z + E_{el} \\
    &= -\tfrac{1}{2}\sum_{ij}J (\boldsymbol{r}_{ij})\boldsymbol{S}_i\cdot\boldsymbol{S}_j \\
    &\quad - \tfrac{1}{2}J_{\parallel}\sum_{ij\in NN}\tfrac{1}{|\boldsymbol{r_{ij}}|^2}(\boldsymbol{S}_i\cdot\boldsymbol{r}_{ij})(\boldsymbol{r}_{ij}\cdot\boldsymbol{S}_j)\\
    &\quad + A_4\sum_{i,\alpha=x,y,z}(S_i^{\alpha})^4 
       - g\mu_B\mu_0\sum_i\boldsymbol{H}\cdot\boldsymbol{S}_i \\
    &\quad + \tfrac{V}{2}\sum_{\alpha\alpha'\beta\beta'}c^{\alpha\alpha'\beta\beta'}\varepsilon_{\alpha\alpha'}\varepsilon_{\beta\beta'},
\end{split}
\end{equation}
with the coupling constants according with \cite{gao_2020} $J_1$ = 0.053\,meV, $J_2$ = -0.079\,meV, $J_3$ = -0.015\,meV, and the anisotropic coupling strength $J_{\parallel}$ = 0.005\,meV and the single ion anisotropy constant $A_4$ = 0.14\,$\mu$eV. $\boldsymbol{H}$ is the external magnetic field, $\boldsymbol{S}_i$ is the spin of ion i in units of $\hbar$, and $\boldsymbol{r}_{ij}$ is the connection vector between ions i and j. 
We derive a set of equations for the strain components $\varepsilon_{\beta\beta'}$ by including the strain dependence of the exchange interactions and minimizing the free energy with respect to  $\varepsilon_{\beta\beta'}$. The result is
\begin{equation}
\begin{aligned}
    \sum_{\beta,\beta'=1...3}c^{\alpha\alpha'\beta\beta'}\varepsilon_{\beta\beta'} 
    =\hphantom{xxxxxxxxxxxxxxxxxxxxx}\\[-0.3em] 
    \hphantom{xxx}\frac{1}{4V}\sum_{ij}\left(r_{\alpha}\frac{\partial J (\boldsymbol{r}_{ij})}{\partial r_{\alpha'}} + r_{\alpha'}\frac{\partial J (\boldsymbol{r}_{ij})}{\partial r_{\alpha}}\right)\langle\boldsymbol{S}^i\cdot\boldsymbol{S}^j\rangle,
\end{aligned}
    \label{strain}
\end{equation}
which, in general, represents six independent equations. $\langle\boldsymbol{S}^i\cdot\boldsymbol{S}^j\rangle$ is the scalar
product of neighboring spin moments. 
The non-zero components of the symmetric elastic-constants tensor $c^{\alpha\alpha'\beta\beta'}$ were calculated by the use of an empirical Born von Karmán model as
\begin{equation}
\begin{split}
    c^{\alpha\alpha\alpha\alpha} &= 93\,\,\mathrm{GPa}\\
    c^{\alpha\alpha\beta\beta} = c^{\alpha\beta\alpha\beta} &= 29\,\,\mathrm{GPa},
\end{split}
\end{equation} 
with new components $\alpha,\beta$ = 1, 2, 3 and $\alpha\neq\beta$.
In order to calculate $\partial J(\boldsymbol{r}_{ij})/{\partial \boldsymbol{r}_{ij}^{\alpha'}}$ we introduced a parametrization \cite{kaneyoshi73} for $J(\boldsymbol{r_{ij}})$ based on $J_1$, $J_2$, and $J_3$ given by
\begin{equation}
  J(\boldsymbol{r}_{ij}) = A\left(r_0^2 - r_0^4\right) e^{-\gamma  r_0^2},
\end{equation}
with $r_0 = |\boldsymbol{r}_{ij}|/R$ and obtained the corresponding constants $A$ = 4.864 meV, $R$ = 4.740
{\AA}, and $\gamma$ = 2.450. 

\begin{figure}[t]
\includegraphics[width=0.85\columnwidth]{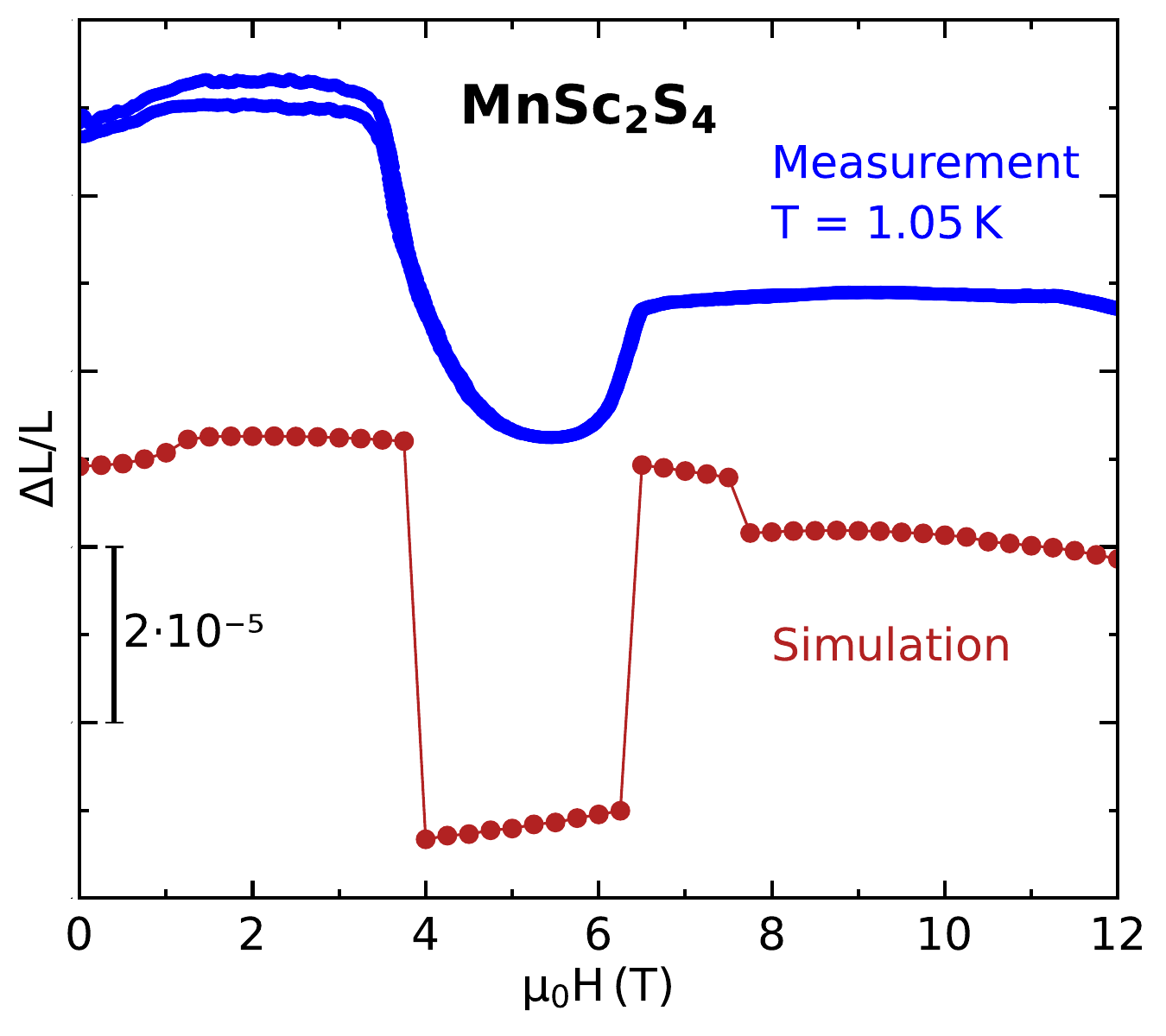}
\caption{Simulated magnetostriction (red) of MnSc$_2$S$_4$ using
\textsc{McPhase} with fields applied along the [111] direction compared
to the measured magnetostriction at 1.05 K (blue) with subtracted
phononic background (magnetostriction data measured at 2.75 K, see Fig.\ \ref{MSS_dil}).}
\label{simstrik}
\end{figure}

We compare the data of simulation and experiment of the magnetic contribution
to the magnetostriction in Fig.\ \ref{simstrik}. The qualitative
agreement between theory and experiment is striking. The sequence of
transitions is the same for the calculated propagation vectors. At zero
field and very low temperatures, the simulation results in a single-\textit{\textbf{q}} state with propagation
vector (3/4, 3/4, 0), corresponding to a helical structure with nonsymmetric
domains \cite{HbGRA}. A transition to a 3\textit{\textbf{q}} state appears
at about 4 T, corresponding to the skyrmion lattice. This transition shows
a strong magnetoelastic response. Above about 6 T, we find another
single-\textit{\textbf{q}} structure with the same propagation as the
former. Based on the calculation, both single-\textit{\textbf{q}} phases
are similar, but our measurements show a clear difference in the lattice
expansion, indicating a different, most likely fan-like single-\textit{\textbf{q}} phase. Finally, the simulation shows a step in $\Delta L/L$
around 7.5 T, which fits to the transition $\Romanbar{5}$. The neutron-scattering intensities of the simulated propagations suggest that this new phase has a 2\textit{\textbf{q}} structure.

We can explain the strong lattice contraction in the 3\textit{\textbf{q}}
skyrmion phase directly from the linear dependence of strain and
nearest-neighbor correlation $\langle\boldsymbol{S}^i\cdot \boldsymbol{S}^j
\rangle$ in Eq.\ (\ref{strain}). In the 3\textit{\textbf{q}} state, each
magnetic moment is aligned almost parallel to the neighboring moment, only
rotated by a small twist angle. In the single-\textit{\textbf{q}} state,
on the other hand, each magnetic moment has two nearly parallel neighbors
in the propagation direction and four neighbors with antiparallel
contributions, due to the overall antiferromagnetic coupling. This leads
to a jump in the nearest-neighbor correlation at the transition from the
single-\textit{\textbf{q}} to the 3\textit{\textbf{q}} state, which directly
influences the magnetostriction, as seen in our experiments and simulations.
The slightly smaller $\Delta L/L$ for the single-\textit{\textbf{q}} phase
above transition \Romanbar{4} compared to the ground state is in line with
the observed fan state at higher fields \cite{gao_2020} according to the
explanation by nearest-neighbor correlations. This small difference occurs
because the moments of neighboring Mn ions are aligned less antiparallel
to each other in the fan state compared to the helical ground state.

\subsection{\texorpdfstring{MnSc$_2$Se$_4$}{MnSc2Se4}}
In contrast to MnSc$_2$S$_4$ we could not find long-range magnetic order in MnSc$_2$Se$_4$, which contradicts the
neutron-scattering results on powder samples \cite{Guratinder_2022}.
This may be caused by internal stresses in single crystals compared
to the powder samples. Here, we can show that an explanation that
allows a helical ground state is also possible. The broad maximum in the
specific heat may indicate the lack of a distinct energy scale. This is
typical for systems with disorder or randomness, which may be caused by
random site exchanges of the cations (antisite defects), which is often
the case in Se systems \cite{Houck2019, WEI2005, Canepa_2017}.
Introducing randomness in this way is used since more than 50
years \cite{BENABRAHAM19689}.
These site exchanges between Mn and Sc (local disorder) are energetically favorable also in MnSc$_2$Se$_4$ crystals.
Due to different internal distortions, the distance of neighboring
Mn sites might vary. In consequence, the exchange interactions vary
as well. All together, larger bond lengths and antisite-defects are connected to each other and cannot be separated experimentally.

However, in our simulation we can tune the amount of defects, which allows us to find hints which effect causes the different behavior.
In particular, we performed a calculation of the specific heat based on the exchange
interactions $J_1$, $J_2$, and $J_3$. We assumed a Gaussian variation of
the exchange coupling with mean values $J_i$ ($i=1,2,3$) and standard
deviation $W$ for each exchange. This approach to model the antisite defects is not possible with \textsc{McPhase}.
Instead, we used the Metropolis algorithm to perform Monte Carlo calculations of a classical Heisenberg model with interactions up to the third nearest neighbor.
Our simulations are similar to former calculations \cite{Miyata_2020,Gen_2022}. We set the mean coupling
constants to the optimal values obtained from fits to the magnon dispersion
($J_1$ = 0.050\,meV, $J_2$ = -0.062\,meV, $J_3$ = -0.008\,meV)
\cite{Guratinder_2022}. We determined the specific heat from the energy
variance.

\begin{figure}[t]
  \centering
  \includegraphics[width=0.98\columnwidth]{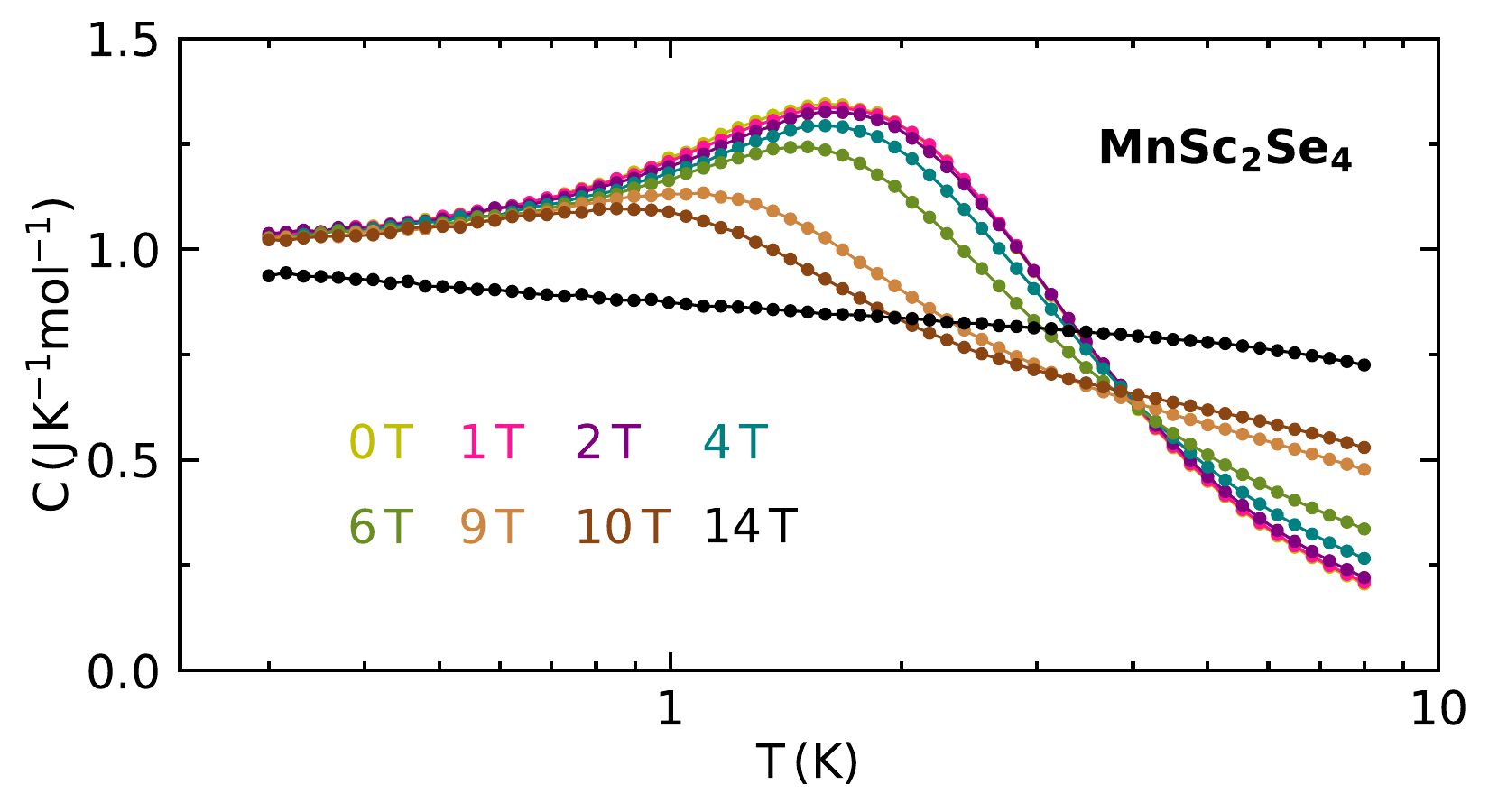}
  \caption{Magnetic specific heat of MnSc$_2$Se$_4$ obtained by Monte
  Carlo simulations of a $J_1$-$J_2$-$J_3$ model with a bond randomness
  parameter $W$ with fields applied along [111].}
  \label{fig:simSH}
\end{figure}

At zero magnetic field, the pure system with zero standard deviation
($W=0$) shows a sharp divergent peak in the specific heat at a first-order
phase transition with $T_N \approx 2$ K. This would evidence long-range order.
Increasing randomness of the exchange broadens the specific-heat peak and
the finite-temperature phase transition disappears above a threshold of
$W$ between 0.04 and 0.08 K. We chose $W=0.18$ K to best reproduce the
peak width of the experimentally measured specific heat (Fig.\ \ref{spech}).
We show the calculated data for several magnetic fields in Fig.\ \ref{fig:simSH}.
Note that the measured specific heat in Fig.\ref{spech} still includes the contribution from
phonons, contrary to the Monte Carlo results. The calculated field and
temperature dependence of the specific heat qualitatively agrees with the
experimental data. This agreement indicates that bond randomness may be significant
in MnSc$_2$Se$_4$ and avoids an overall helical ground state. Increased bond lengths alone would allow a sharp transition to long range magnetic order, based on our simulation. We cannot decide, if magnetic structures, like the SkL would also occur without bond-randomness. But this is quite speculative, since there is no possibility to exclude antisite-defects in MnSc$_2$Se$_4$.

\section{Summary}
We investigated the spinel compounds MnSc$_2X_4$ ($X$ = S, Se) by using
ultrasound and dilatometry experiments, supported by further thermodynamic
methods as well as model calculations. For single-crystalline MnSc$_2$S$_4$,
we constructed a rich ($H,T$) phase diagram based on the anomalies found
in the ultrasound and dilatometry data. At zero field, our data indicate
the existence of three magnetic phases in line with earlier results
\cite{gao_2017, gao_2020}. A commensurable single-\textit{\textbf{q}} helical ground state at lowest temperature has been proposed, which is followed by an intermediate incommensurable state and a collinear state prior to the transition into paramagnetism. 
In magnetic fields applied along [111],
we observe clear anomalies evidencing the phase transitions to the
3\textit{\textbf{q}} skyrmion phase seen in neutron scattering \cite{gao_2020}.
This skyrmion phase is restricted to a roughly 2~T broad region. It turns
out that the skyrmions couple strongly to the lattice and induce pronounced
effects in the magnetoelastic properties.
This is similar to other compounds, \textit{e.g.} the 4f-magnet Gd$_2$PdSi$_3$ \cite{Spachmann_2021} or FeGe \cite{Mito_24}, including volume variations.
We further observe two new phase-transition lines in MnSc$_2$S$_4$,
one of which may separate a 2\textit{\textbf{q}} phase.

For MnSc$_2$Se$_4$, on the other hand, we could not detect any sharp
phase transition. Only a broad peak in the specific heat appears at about 2 K.
This may indicate a spin-liquid state. However, assuming some randomness
in the lattice, due to larger bond lengths, our Monte Carlo calculations can explain the broad
specific-heat peak and still support the possibility of a helical ground state. Nevertheless,
unlike MnSc$_2$S$_4$, MnSc$_2$Se$_4$ does not contain a skyrmionic state.

Further intriguing changes in the magnetic properties might occur
with complete or partial substitutions on the Sc or $X$ sites, which
may clarify further details about the Mn spinels.

\begin{acknowledgements}
We acknowledge support from Deutsche Forschungsgemeinschaft (DFG)
through SFB 1143 (Project-ID 247310070), the W\"urzburg-Dresden Cluster
of Excellence on Complexity and Topology in Quantum Matter---$ct.qmat$
(EXC 2147, Project No.\ 390858490), and TRR 360 (Project No.\ 492547816). We as well acknowledge the support
of the HLD at HZDR, member of European Magnetic Field Laboratory (EMFL).
We thank Oksana Zaharko for fruitful discussions. L.P. and V.T thank for support by the project ANCD (code 011201, Moldova). H.S. acknowledges support from
JSPS KAKENHI Grant Nos. JP22K03508 and JP24H01609. Monte Carlo calculations
were performed using computational resources of the Supercomputer Center
at the Institute for Solid State Physics, the University of Tokyo.
\end{acknowledgements}

\section*{Data availability}

The data that support the findings of this article are openly available Ref. \cite{spinel_TUD}.

\bibliography{spinel_cites}


\end{document}